\documentclass[12pt]{iopart}
\usepackage{iopams}
\usepackage{graphicx}
\usepackage{cite}

\def\bt{\textbf{t}}
\def\bs{\textbf{s}}

\newtheorem{theo}{Theorem}
\newtheorem{defi}{Definition}

\begin{document}
\title{Spectral curves in gauge/string
dualities: integrability, singular sectors and regularization. }
\author{ Boris Konopelchenko$^1$, Luis Mart\'{\i}nez Alonso$^2$  \\ and Elena Medina$^3$}
\address{$^1$ Dipartimento di Matematica e Fisica  ``Ennio De Giorgi'',
                        Universit\'a del Salento and Sezione INFN,
                        7310 Lecce, Italy}
\address{$^2$ Departamento de F\'{\i}sica Te\'orica II,
                        Facultad de Ciencias F\'{\i}sicas,
                        Universidad Complutense,
                        28040 Madrid, Spain}
\address{$^3$ Departamento de Matem\'aticas,
                        Facultad de Ciencias,
                        Universidad de C\'adiz,
                        11510 Puerto Real, C\'adiz, Spain}
\begin{abstract}
We study the moduli space of the  spectral curves $y^2=W'(z)^2+f(z)$ which characterize  the   vacua  of  $\mathcal{N}=1$ $U(n)$ supersymmetric
gauge theories with an adjoint Higgs field  and a polynomial  tree level potential $W(z)$.
 The  integrable structure of the Whitham equations  is used  to
determine the spectral curves from  their moduli.  An alternative   characterization of the spectral curves  in terms of  critical points
of a family of  polynomial solutions $\mathbb{W}$ to Euler-Poisson-Darboux equations is provided. The equations for these critical points
are a generalization of  the planar limit equations for  one-cut random matrix models. Moreover,  singular spectral curves with higher
order branch points  turn out to be  described by  degenerate critical points of $\mathbb{W}$.  As a consequence we propose a  multiple scaling
limit method of regularization and show that, in the simplest cases, it leads to the Painlev\`{e}-I equation and its multi-component generalizations.

\end{abstract}
\vspace{1cm}
\emph{J. Phys. A: Math. Theor. 46 (2013)}
\maketitle
%

\section{Introduction}

In a series of seminal papers ~~\cite{CA01,DI02, DI022}  Cachazo, Dijkgraaf, Intriligator and Vafa  proved
that a class of  hyperelliptic curves of the form
\begin{equation}\label{sp0}
y^{2}=W'(z)^{ 2}+f(z),
\end{equation}
where $W(z)$ and $f(z)$ are polynomials such that ${\rm deg} f={\rm deg} W-2$,
are essential to study   the low energy dynamics of   pure $\mathcal{N}=2$   $U(n)$ supersymmetric gauge theories  deformed to $\mathcal{N}=1$ by  a tree level potential
${\rm Tr}\, W(\Phi)$ for the Higgs field $\Phi$.
 These curves  emerge from the duality
between supersymmetric  gauge theories and  string theories  on deformed 3-dimensional Calabi-Yau (CY) manifolds defined by the equation
\begin{equation}\label{cym}
W'(z)^2+f(z)+u^{2}+v^{2}+w^{2}=0.
\end{equation}
 Here the polynomial $f(z)$ is
 introduced to regularize the singular CY manifold
 \[
 W^{\prime 2}+u^{2}+v^{2}+w^{2}=0
 \]
 according to the process called \emph{Geometric Transition } ~\cite{CA01}.  The curves (\ref{sp0}) can be also
 generated from a special class of  Seiberg-Witten curves ~\cite{SE94}
of pure  $\mathcal{N}=2$  gauge theories which satisfy  a certain  factorization property ~\cite{CA02b}.

  Curves of the form (\ref{sp0}) are   relevant  in the study of the asymptotic eigenvalue distribution as $n\rightarrow
\infty $ of  matrix models with partition functions
\begin{equation}
Z_{n}=\int e^{-n\,\mathrm{Tr}\,W(M)} \rmd M\,,  \label{mm0}
\end{equation}
where $n$ is the matrix dimension.  The best known matrix  model is the  Hermitian model  ~\cite{DE99,BL08},  but more general models based on sets of matrices such that the eigenvalues are constrained to
lie on some fixed path $\Gamma$ in the complex plane are used in the analysis of gauge/string dualities ~\cite{AL10,LA03,FE04,BI05}
 .  The curves (\ref{sp0}) also  arise  in the study of the asymptotic
zero distribution as $n\rightarrow \infty $ of monic orthogonal polynomials $
P_{n}(z)=z^{n}+\cdots $, verifying the conditions  ~\cite{GO87,GO89,BL99,BE06,BE09,BE11,MA11,RA12}
\begin{equation}
\int_{\Gamma }P_{n}(z)\,z^{k}\,e^{-n\,W(z)}\,\mathrm{d}z=0,\quad k=0,\ldots
,n-1,  \label{pol1}
\end{equation}
on a certain path $\Gamma$ in the complex plane.

The curves (\ref{sp0})  have an even number $2q \,  (1\leq q \leq N)$ of branch points
given by the odd-order roots
$
\boldsymbol\beta=(\beta_1,\ldots,\beta_{2q}),
$
of the polynomial $y(z)^2$. In the applications to gauge/string dualities
the curves (\ref{sp0})  are characterized by imposing  $q$ \emph{ period relations } of the form
\begin{equation}\label{toof}
\oint_{A _{j}}y(z)\rmd z=-4\,\pi \,\rmi\,s_{j},\quad j=1,\ldots ,q,
\end{equation}
where $\bs=(s_1,\ldots,s_q)$ are $q$ given  arbitrary  complex numbers called \emph{'t~Hooft parameters}. Here
 $y(z)$ is the branch of (\ref{sp0}) with asymptotic behavior
        \begin{equation}
            \label{yin}
            y(z) = W'(z) + \mathcal{O}(z^{-1}),\quad  z\rightarrow\infty,
        \end{equation}
and $A_j$ are counterclockwise cycles around $q$ disjoint oriented curves  (\emph{cuts})  $\gamma _{j}$ with endpoints $(\beta_{2j-1},\beta _{2j})$.

The present work is a natural continuation of our study of phase transitions in random matrix models  ~\cite{AL10} and in the space of vacua in supersymmetric  gauge theories ~\cite{AL13}.  Our main now is to study the structure  of the moduli space of the curves (\ref{sp0})  which satisfy a system of period relations of the form  (\ref{toof}).  We assume that
\begin{equation}\label{sp1}
 W(z)=\frac{z^{N+1}}{N+1}+\sum_{n=1}^{N}g_n\,z^{n}, \quad  f(z)=\sum_{k=1}^{N}z^{N-k}\,t_{k},
 \end{equation}
where the coefficients $g_k$ are fixed while the coefficients $t_k$ are considered as  arbitrary  varying complex  parameters.
Henceforth we will refer to (\ref{sp0}) as \emph{spectral curves} and to the coefficients  ${\bf t}=(t_1,\ldots,t_N)$  as the
\emph{deformation parameters}.  Our main interest is to apply the theory of integrable systems to describe the analyticity
properties of the roots of $y(z)^2$ as functions of  either  the 't Hooft parameters or the deformation parameters.

 From the point of view of the underlying $U(n)$ gauge theory,  $q$-cut spectral curves are associated  to classical
 vacua such that the $n$  eigenvalues of the Higgs field $\Phi $ distribute among $q$ critical points $a_j$ of $W(z)$
 with multiplicities $n_j$ where $n_1+\ldots +n_q=n$. On these vacua the gauge group is spontaneously broken into $q$ factors
\[
U(n)\rightarrow U(n_1)\times U(n_2)\times \cdots \times U(n_q).
\]
The role of
  spectral curves  in  the low energy dynamics of the $U(n)$ gauge theory  is of central importance since they determine the
  \emph{prepotential functional}
 (see ~\cite{CA01},~\cite{DI02} and ~\cite
{DI022})
\begin{equation}\label{free}
\mathcal{F}=\int_{\gamma }\rho (z) W(z)|\rmd z|-\frac{1}{2}\int_{\gamma
}\int_{\gamma }\rho (z)\,\rho (z')\log
(z-z')^2 |\rmd z||\rmd z'|,
\end{equation}
which is the key object to obtain the vacuum expectation values
of many important operators of the quantum theory. Here $\gamma$ denotes the union of all the cuts
\[
\gamma=\gamma_1\cup\cdots \cup \gamma_q,
\]
and the function $\rho (z)$  (\emph{the spectral density}) is defined by
\begin{equation}\label{den}
\rho (z)|{\rmd} z|=\frac{y(z_+){\rm d} z}{2\pi \rmi}=-\frac{y(z_-){\rmd}z}{2\pi \rmi}
,\quad z\in \gamma.
\end{equation}
The subscripts $\pm$ in  $z_{\pm}$ indicate the limits of functions from the left  and right of the oriented cuts.
The `t Hooft parameters can be considered as \emph{partial charges}  with respect to the spectral density since  we may rewrite (\ref{toof}) as
\begin{equation}
    \label{den1}
    \int_{\gamma_j}  \rho(z) | \rmd z| =  s_j.
\end{equation}
The spectral density obviously depends  on the choice of the cuts $\gamma_j$.  However, as it is known in the study of gauge/string dualities ~\cite{CA01}-~\cite{DI022}, the prepotential as a function of the 't Hooft parameters  is determined by the periods of the differential $y(z) {\rm d} z$ on a certain set of non-compact cycles in (\ref{sp0}) (\emph{special geometry relations)}. As a consequence  $\mathcal{F}(\bs)$ is invariant under deformation of the cuts in the same homology classes of $\mathbb{C}\setminus \{\beta_i\}_{i=1}^{2q}$.

In the present  work we  introduce a decomposition of  the moduli space of spectral curves onto classes  in terms of sets
$\mathcal{M}_q$  of spectral curves  with at most $2q$ branch points.  Each class $\mathcal{M}
_q$ is naturally stratified into a regular and a singular sector.  The regular sectors contain the spectral curves
which generically are determined by the constraints (\ref{toof}),  so that they
depend on  $q$ moduli which may be either  the 't Hooft parameters  or  a subset of $q$ deformation parameters
.  In  the language  of  supersymmetric   gauge theory ~\cite{FE03d}, the regular sectors represent  the different
quantum  phases  of the vacua of the  gauge model and,
consequently,  the singular sectors describe the regions of phase
transitions ~\cite{AL13}.

   We revisit  the theory of Whitham equations for spectral curves   ~\cite{KR04,ch03,ch03b}  to  find   the  differential equations
which characterize the  classes   $\mathcal{M}_q$  and  to determine the spectral curves from systems of implicit equations
involving the branch points of $y(z)^2$ and both the 't Hooft and the deformation parameters. We also give a simple proof of the $\tau$-function character of $\log\mathcal{F}$.
 Furthermore, we provide an alternative way  to  characterize the spectral
curves (\ref{sp0}) from of a \emph{reduced system} of  implicit equations
for  the branch points involving  the $q$ deformation parameters $(t_1,\ldots,t_q)$ only.
The reduced system  represents  the  critical points of certain
polynomial solutions $\mathbb{W}$ of  Euler-Poisson-Darboux equations ~\cite{DAR},  and  it turns out to be a  generalization
for multicut spectral curves  of the planar limit  equations  of one-cut ($q=1$) random matrix models  ~\cite{DI95}-~\cite{KO11}.
The main advantage  of the use of reduced systems is that they  drastically simplify the analysis of
singular sectors where the gradient catastrophe for the branch points and, consequently,   the breaking of analyticity for the
spectral curve and the free energy (\ref{free}) happens. In this way we show how singular  spectral curves corresponding
to the emergence of higher order branch points can be described in terms of simple classes of degenerate  critical
points of the solutions $\mathbb{W}$ of the  Euler-Poisson-Darboux equations. Moreover, the  reduced systems leads naturally to
the multiple scaling limit method to regularize the critical behavior of the branch points.  Thus we propose a method of
regularization of the gradient catastrophes  based on
 substituting  the  critical point equations for the functions $\mathbb{W}$  by the  Euler-Lagrange equations  for  a  density functional $\mathbb{W}
^{reg}$  obtained by
adding appropriate terms with derivatives to $\mathbb{W}$.
In  the simplest cases the
corresponding regularizing differential equations are  the Painlev\`{e}-I equation and its multi-component extensions.

The paper is organised  as follows. In section 2 we analyze the moduli space
of the  spectral curves and define the  classes $\mathcal{M}_q$  as well as their
  regular and singular sectors. We  present a method to determine spectral curves from the set of 't Hooft parameters and introduce the  notions of spectral density and the prepotential functional.  Then we  analyze  the integrable structure  supplied by  Whitham equations  with respect to  't Hooft parameters.   Section 3 deals with the analysis of the moduli space of spectral curves  in terms of  deformation parameters
     and with the  study of the analyticity properties  of the roots of $y(z)^2$  as functions of the deformation parameters.
     We  illustrate our theoretical scheme with
 a complete description of the moduli space of  spectral curves for the Gaussian and cubic models
 \[
W(z)=\frac{z^2}{2},\; \frac{z^3}{3}-g\,z.
\]
Section 4 presents the method of reduced systems  based on solutions of  Euler-Poisson-Darboux equations to determine spectral curves.
We also show  the connection between singular spectral curves and degenerate solutions of reduced systems.  Sections 5 and 6 describe a
simple implementation of the multiple scaling limit to regularize gradient catastrophes of the branch point functions $\boldsymbol \beta(\bt)$.
Moreover,  the natural emergence of the Painlev\`{e}-I equation
is demonstrated.
\section{The moduli space of spectral curves }

We may distinguish different classes in the  set of spectral curves (\ref{sp0}) by looking at the multiplicities  $[m_{i_1},m_{i_2},\ldots], \,(1\leq i_1<i_2<\ldots)$ of the roots of $y(z)^2$,
where $m_i$ is the number of roots with multiplicity equal to $i$.
The polynomial  $y(z)^2$ is of order $2N$,  so that it has an even number $2q$ of roots  $\beta_i$ with odd multiplicities
(branch points of $y(z)$) and a certain number $p$ of roots $\alpha_l$  with even multiplicities.

 As we said in the introduction,  the spectral curves (\ref{sp0}) arising  in the study of gauge/string
dualities are determined imposing  $q$ period conditions (\ref{toof}).
Thus we will henceforth concentrate on  families of spectral curves such that
  $y(z)^2$ is of the form
  \begin{equation}\label{phys}
y^2(z)=\prod_{l=1}^p (z-\alpha_l)^{2}\,\prod_{i=1}^{2\,q}(z-\beta_i),\quad p+q=N.
\end{equation}
 Indeed these families depend generically on $q$ free parameters since  equating (\ref{sp0}) and (\ref{phys}),  then identifying  coefficients of powers of $z$  leads to a system of $2\,N$ equations for determining the $2\,N+q$ variables $(\alpha_l,\beta_i,t_k)$.  We will see that
 the moduli space of the spectral curves (\ref{sp0}) which satisfy  (\ref{toof}) and are of the form (\ref{phys})  can be described in terms of the set  of  't Hooft parameters  $\bs=(s_1,\ldots,s_q)$ .

\subsection{Regular and singular sectors of spectral curves}
In order to deal with the structures of singularities we will consider degenerations of the factorization (\ref{phys}) in which some roots coalesce.
This motivates the following definition:

\begin{defi}
Given $q=0,1,\dots,N$ we denote by $\mathcal{M}_q$ the set of all the  spectral curves which have $2q$ branch points  at most.
\end{defi}

The class  $\mathcal{M}_q$ represents the set of all spectral curves which admit  a factorization  of the form (\ref{phys}) where
some of the roots $(\alpha_l,\beta_i)$ may coincide.
 Obviously we have

 \begin{equation}\label{part2}
 \mathcal{M}_0\subset  \mathcal{M}_1\cdots \subset  \mathcal{M}_N.
 \end{equation}

\begin{defi}
We define the regular sector   ${\rm reg}\mathcal{M}_q$  as the subset of the spectral curves in  $\mathcal{M}_q$
which have  all the roots $(\alpha _{l},\beta _{i})$ distinct. We define the singular sector of $\mathcal{M}_q$
by ${\rm sing} \mathcal{M}_q=\mathcal{M}_q\setminus {\rm reg} \mathcal{M}_q$.

\end{defi}

  The regular sector  ${\rm reg}\mathcal{M}_q$ represents the set of  spectral curves such that   $y(z)^2$  factorizes in the form (\ref{phys}) with distinct roots $(\alpha_l,\beta_i)$. The spectral curves
  in the singular sector ${\rm sing} \mathcal{M}_q$ admit a factorization of the form (\ref{phys})  but such that some
  of the roots $(\alpha _{l},\beta _{i})$  coincide.
It is  clear that $\mathcal{M}_{q'}\subset {\rm sing}\mathcal{M}_q$ for $q'<q$.

\subsection{Determination of spectral curves from 't Hooft parameters}

To determine spectral curves of the form (\ref{phys}) which satisfy  the period relations (\ref{toof})  we notice that  the condition (\ref{yin}) implies
\begin{equation}
    \label{ache}
    \prod_{l=1}^p (z-\alpha_l)=\left( \frac{W'(z)}{y_0(z)} \right)_{\oplus},
\end{equation}
where $y_0(z)$ is the function
\begin{equation}
  \label{1.0}
  y_0(z)=\sqrt{\prod_{i=1}^{2q}(z-\beta_{i})},
\end{equation}
such that $y_0(z)\sim z^q$ as $z\rightarrow\infty$ and $\oplus$ stands for the sum of the nonnegative powers of the Laurent series of $W'(z)/y_0(z)$ at infinity. Then from (\ref{ache}) we may determine  the double roots $\alpha_l$ as functions of  the simple roots $\beta_i$. Hence
the function $y^2(z)$ is completely determined by the set of simple roots $\beta_i$.

The simple roots $\beta_i$  can be determined as follows.  We first substitute
\[
y(z)=\left( \frac{W'(z)}{y_0(z)} \right)_{\oplus}y_0(z)
\]
 in (\ref{sp0}) and identify  the coefficients of
$1,z,\cdots, z^{N+q-1}$ in both members of (\ref{sp0}). The remaining coefficients do not give independent
relations because as $z\rightarrow \infty.$
\begin{equation*}
    y(z)^2-W'(z)^2
     =  \Big(\frac{W'(z)}{y_0(z)}\Big)_{\ominus}
             \Big[\Big(\frac{W'(z)}{y_0(z)}\Big)_{\ominus} y_0(z)- 2 W'(z)\Big]\,y_0(z)
         =  \mathcal{O}(z^{N+
        q-1}).
\end{equation*}
Thus we find $N+q$ equations for the $N+2q$ variables $(\beta_i,t_k)$.  The  required $q$  additional independent relations are provided by the period conditions
\begin{equation}\label{toof1}
\oint_{A _{j}}y(z)\rmd z=-4\,\pi \,\rmi\,s_{j},\quad j=1,\ldots ,q,
\end{equation}
where $\bs=(s_1,\ldots,s_q)$ are the \emph{'t~Hooft parameters} and $A_j$ are counterclockwise cycles around fixed cuts $\gamma_j$ connecting the pairs $(\beta
_{2j-1},\beta _{2j})$ of branch points.   It is clear that the period conditions only depend  on the homology classes of the cuts in $\mathbb{C}\setminus \{\beta_i\}_{i=1}^{2q}$.
Hence the natural choice is to take each cut $\gamma_j$ in the homology class of  the straight line segment $[\beta
_{2j-1},\beta _{2j}]$, so that (\ref{toof1}) can be rewritten as
\begin{equation}\label{toof2}
\int_{\beta_{2j-1}}^{\beta_{2j}}y(z_+)\rmd z=2\,\pi \,\rmi\,s_{j},\quad j=1,\ldots ,q.
\end{equation}
In this way we have a method for determining  the $N+2q$ variables $(\beta_i,t_k)$ as functions of the  't Hooft parameters  by means of a system of $N+2q$ implicit equations.
Moreover,  from (\ref{toof2}) it is clear that for $q>1$  the set of solutions  provided by  this  method   depends on how we split the $2q$ simple roots into pairs to define the cuts.  In the applications  to gauge/string dualities  each  of these solutions  characterizes a different  \emph{classical limit} $\bs\rightarrow \bf{0}$. In particular if the deformation parameters of a solution  vanish  as  $\bs\rightarrow \bf{0}$
  then the polynomial $f$ vanishes and  the spectral curve reduces to
$
y^{2}=W'(z)^2.
$
Thus, the $2q$ simple roots  of the spectral curve corresponding  to this solution   represent a regularization of  $q$ double roots  of the undeformed curve.

In general, as we will see below in the analysis of the cubic model,  this method for determining the cut endpoints leads to several families of solutions  $(\beta_i(\bs),t_k(\bs))$ which determine several families
of spectral curves in ${\rm reg}\mathcal{M}_q$.


\subsection{The spectral density}

Let us consider  a spectral curve in the regular sector ${\rm reg}\mathcal{M}_q$ and  take  a semi-infinite oriented regular path $\Gamma$ (see figure 1) containing all the cuts $\gamma_j$. Henceforth we will assume that  $\Gamma$ admits a  parametrization $z(t)=(x(t),y(t))$ such that at least one of the components is a strictly  monotone function of  $t$.
For each $z'$ in $\Gamma$ let us denote by $\Gamma_{z'}$ the semi-infinite arc of $\Gamma$ ending at $z'$ and  define
\begin{equation}
    \label{logbb}
    \log(z-z') = \log |z-z'| + \int_{\Gamma_{z,z'}} \frac{\rmd u}{u-z'},
    \quad z'\in \Gamma,
    \quad z\in \mathbb{C}\setminus \Gamma_{z'},
\end{equation}
where $\Gamma_{z,z'}$ is any path in $\mathbb{C}\setminus \Gamma_{z'}$ connecting $z'+|z-z'|$
to $z$. It is clear that   $\log(z-z')$
depends analytically
of $z$ in $\mathbb{C}\setminus \Gamma_{z'}$.
 For example if $\Gamma$ is a real interval of the form $(-\infty, x_0]$  then (\ref{logbb}) determines the principal branch of $\log(z-z')$.
\begin{figure}
    \begin{center}
        \includegraphics[width=8cm]{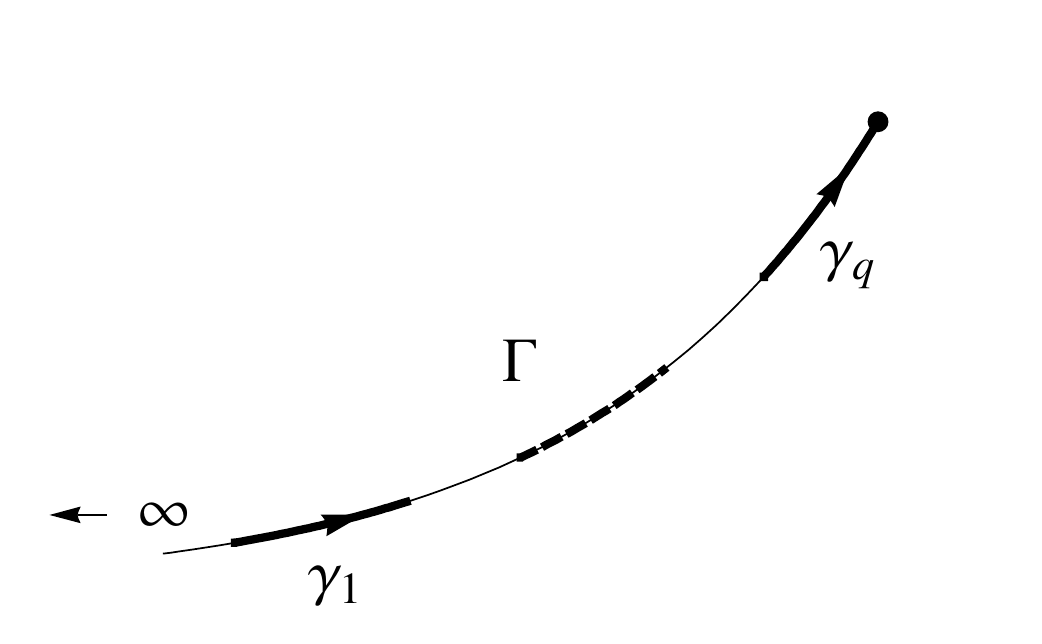}
    \end{center}
    \caption{The semi-infinite path $\Gamma$ ending at $\beta_{2q}$.
    \label{fig1}}
\end{figure}
Let us now introduce the function
\begin{equation}
    \label{ge}
    g(z) = \int_{\gamma}\log(z-z') \rho(z') |\rmd z'|,\quad \gamma=\gamma_1\cup\cdots \cup \gamma_q,
\end{equation}
where $\rho(z)$ is the spectral density
\begin{equation}\label{den1}
\rho (z)|{\rmd} z|=\frac{y(z_+){\rm d} z}{2\pi \rmi}=-\frac{y(z_-){\rmd}z}{2\pi \rmi}
,\quad z\in \gamma.
\end{equation}
Note that both $y(z)-W'(z)$ and $-2g'(z)$
are analytic in $\mathbb{C}\setminus\gamma$,  vanish as $z\to\infty$ and, according to~(\ref{den}),
have the same jump on $\gamma$. Therefore we obtain the identity
\begin{equation}
    \label{y}
    y(z) = W'(z)-2 g'(z)=W'(z)-2\, \int_{\gamma}\frac{\rho(z')}{z-z'} |\rmd z'|,
\end{equation}
which determines the spectral curve from the spectral density. It also implies the asymptotic behavior
\begin{equation}\label{asb}
y(z)=W'(z)-2\,g'(z)=W'(z)-\frac{2s}{z}+\mathcal{O}(z^{-2}),\quad \mbox{as $z\rightarrow \infty$},
\end{equation}
where $s$ denotes the so called \emph{total t'Hooft parameter}
\[
s=s_1+\cdots+s_q.
\]
Moreover, using
\begin{equation}
    \label{rh}
    y(z_+) +y(z_-)=0,\quad z\in \gamma,
\end{equation}
we get
\begin{equation}
    \label{ds}
    W'(z) - (g'(z_+)+g'(z_-)) = 0,\quad z\in\gamma.
 \end{equation}
This equation means that $W(z)-(g(z_+)+g(z_-))$ is constant on each connected piece
of $\gamma$ or, equivalently, that there are (not necessarily equal) complex numbers $l_i$ such that
\begin{equation}
    \label{s1}
     W(z)-\left(g(z_+)+g(z_-)\right)=l_i,\quad  \forall z\in \gamma_i.
\end{equation}

The identities (\ref{s1}) are essential  for the applications of spectral curves. There are two particularly important examples:
\begin{itemize}
\item The spectral curves arising in the large $n$-limit of Hermitian matrix models (\ref{mm0}) lead to  normalized spectral densities $\rho(z)$ which represent the asymptotic eigenvalue distribution of the models and are supported on a finite number of disjoint real intervals ~\cite{DE99,BL08}. In the electrostatic interpretation  $\rho(x)$ describes the equilibrium density of a charge distribution on the real line $\mathbb{R}$ under the action of an external real potential $W(x)$. In this case all the parameters $l_i$ (\ref{s1}) are equal and  represent the constant value of the total electrostatic potential on the support $\gamma$ of the equilibrium density $\rho(x)$.
\item The families of orthogonal polynomials on curves $\Gamma$ satisfying the so-called $S$-property (see for example
 ~\cite{GO87}, ~\cite{GO89},
~\cite{MA11} and ~\cite{RA12})  determine spectral curves with associated  spectral densities $\rho(z)$ which represent the asymptotic distribution
of the zeros of the polynomials and are supported on a finite number of finite disjoint curves contained in $\Gamma$. In the electrostatic
interpretation  $\rho(x)$ describes the equilibrium density of a charge distribution on  $\Gamma$ under the action of the external real potential
$\mbox{Re} W$. In this case  all the parameters $l_i$ (\ref{s1}) have the same real part since  this common real part represents the constant value
of the total electrostatic potential on the support $\gamma$ of the equilibrium density $\rho(x)$.

\end{itemize}

The class of spectral curves used in the study of gauge/string dualities generalizes those arising in Hermitian models and families of orthogonal polynomials.

\subsection{The prepotential }

The prepotential  associated to a spectral curve is defined by
 \begin{equation}\label{free2}
\mathcal{F}=\int_{\gamma }\rho (z) W(z)|\rmd z|-\frac{1}{2}\int_{\gamma
}\int_{\gamma }\rho (z)\,\rho (z')\log
(z-z')^2 |\rmd z||\rmd z'|,
\end{equation}
where  we assume that $\log(z-z')^2$ is given by
\begin{equation}
    \label{logq}
    \log(z-z')^2 = \log(z_+-z')+\log(z_--z'),\quad z,z'\in \Gamma.
\end{equation}
The consistence of (\ref{logq}) requires that
\begin{equation}
    \label{loga}
    \log(z_+-z') + \log(z_--z') = \log(z'_+-z)+\log(z'_--z), \;\mbox{$\forall z\neq z'$ in $\Gamma$},
\end{equation}
where the subscripts $\pm$ in  $z_{\pm}$ indicate the limits of functions from the left  and right of the oriented curve $\Gamma$.
It is easy to prove that  the property (\ref{loga}) is satisfied if $\Gamma$ verifies our previous assumption that  at least one of the components  of the parametrization $z=(x(t),y(t))$  is a strictly  monotone function of  $t$.
 For example if $\Gamma$ is a real interval of the form $(-\infty, x_0]$  then (\ref{logbb}) determines the principal branch of $\log(z-z')$
and  (\ref{logq}) yields $\log(z-z')^2=2\log|z-z'|$.

The property  (\ref{loga})  is required to prove that the prepotential
   satisfies the important relations
\begin{equation}
    \label{ele}
    \frac{\partial\mathcal{F}}{\partial s_i} = l_i.
\end{equation}
 Indeed, from  (\ref{s1})- (\ref{free2}),  taking into account  that
$\rho(\beta_j)=0$, we have
\begin{eqnarray*}
        \frac{\partial \mathcal{F}}{\partial s_i}
    & = &\int_{\gamma}\Big(W(z)
    -
\int_{\gamma}\log(z-z')^2 \rho(z')| \rmd z'| \Big)\frac{\partial \rho(z)}{\partial s_i}| \rmd z|\\
        & = &
    \int_{\gamma} \left( W(z) - ( g(z_+)+g(z_-) ) \right) \frac{\partial  \rho(z)}{\partial s_i}| \rmd z| \\\nonumber
    & = &
     \sum_{j=1}^q l_j \int_{\gamma_j} \frac{\partial  \rho(z)}{\partial s_i} | \rmd z| =      \sum_{j=1}^q l_j\frac{\partial s_j}{\partial s_i
     } = l_i.
\end{eqnarray*}

From the definition  (\ref{den1}) of the spectral density it is clear  that  $\rho(z)$ depends on the choice of the cuts $\gamma_j$ connecting the  pairs $(\beta_{2j-1},\beta_{2j})$ of branch  points. However, once the logarithmic branches $\log(z-z')$ are defined, the parameters $l_i$ as well as  the prepotential $\mathcal{F}$  as  functions of the 't Hooft parameters $\bs$  are  invariant under deformations of the cuts  in fixed homology classes in $\mathbb{C}\setminus \{\beta_i\}_{i=1}^{2q}$. To prove this property we follow a method used in ~\cite{FE03d}:  let us take a fixed  $\lambda \in \mathbb{C}\setminus \Gamma$, a fixed $z_j\in \gamma_j$ and two paths $\Gamma_{\lambda,\pm}$ in  $\mathbb{C}\setminus \Gamma$  running from
$z_{j\pm}$  to $\lambda$. Then since $y(z)=W'(z)-2g'(z)$ for  $z\in \mathbb{C}\setminus \Gamma$ and using ~(\ref{s1}) it follows that
\begin{equation}
    \label{rig01}
    \int_{\Gamma_{\lambda,+}}y(z)\rmd z+\int_{\Gamma_{\lambda,-}}y(z)\rmd z=2\left(W(\lambda)-2\,g(\lambda)-l_j\right).
 \end{equation}
 Hence
 \begin{equation}
    \label{rig01}
    l_j=\lim_{\lambda \rightarrow\infty}\Big(W(\lambda)-2 s\,\log\lambda-\frac{1}{2}\int_{\Gamma_{\lambda,+}}y(z)\rmd z-\frac{1}{2}\int_{\Gamma_{\lambda,-}}y(z)\rmd z\Big).
 \end{equation}
From this identity   it is clear that the value of $l_j$ does not change if we deform the cuts  in fixed homology classes in $\mathbb{C}\setminus \{\beta_i\}_{i=1}^{2q}$ keeping the 't Hooft parameters constant. The same statement for the prepotential $\mathcal{F}$ follows at once taking into account that (\ref{toof1}) and (\ref{s1}) imply
\begin{equation}
    \label{pre1}
    \mathcal{F}
    =
    \frac{1}{2}\,\int_{\gamma}\rho(z) W(z)   |\rmd z|   + \frac{1}{2} \sum_{j=1}^q s_j l_j,
\end{equation}
and that
\begin{equation}
  \int_{\gamma}\rho(z) W(z) |\rmd z|=-\frac{1}{4\pi\,i}\sum_{j=1}^q\oint_{A_j} y(z)W(z)\rmd z.
\end{equation}

 \subsection{Integrable structure:  Whitham equations}

Given a  spectral curve (\ref{phys}) in the regular sector  ${\rm reg}\mathcal{M}_q$ we may consider the genus $q-1$ hyperelliptic
Riemann surface  $M$  associated to the reduced curve
\begin{equation}
  \label{1.1}
  y_0^2=\prod_{i=1}^{2q}(z-\beta_{i}),
\end{equation}
which  encodes the complete structure of the spectral curve. We are going to use the theory of Abelian differentials in Riemann surfaces (see Appendix A) to formulate  a system of Whitham differential equations which characterize the cut endpoints as functions $\boldsymbol \beta({\bf s})$ of the 't Hooft parameters. Furthermore, we will derive a system
 of implicit equations which determines  the functions $\boldsymbol \beta({\bf s})$ and exhibits the
 integrability of the  Whitham differential equations.

  Let  $\{A_i, B_i\}_{i=1}^{q-1}$ be  the  homology basis of cycles in $M$ showed in figure 2,  where $A_i$ are counterclockwise
  cycles around the cuts $\gamma_i$ with endpoints $(\beta_{2i-1},\beta_{2i})$. We will  denote
the corresponding periods of a differential $\rmd \omega$ in $M$ by
\begin{equation}
    \label{notp}
    A_i(\rmd \omega) = \oint_{A_i} \rmd \omega,
    \quad
    B_i(\rmd \omega)=\oint_{B_i} \rmd \omega.
 \end{equation}
Let us  introduce  the meromorphic differential
\begin{equation}\label{ds}
\rmd\mathbb{S}=\frac{1}{2}\left(\mathrm{y}(z)+W'(z)\right)\rmd z
\end{equation}
 where $\mathrm{y}(z)$
denotes the function on  the Riemann surface $M$ determined by  two branches of (\ref{sp0})
given by
\begin{equation}\label{two}
y_2(z)=-y_1(z)=y(z).
\end{equation}
\begin{figure}
    \begin{center}
        \includegraphics[width=8cm]{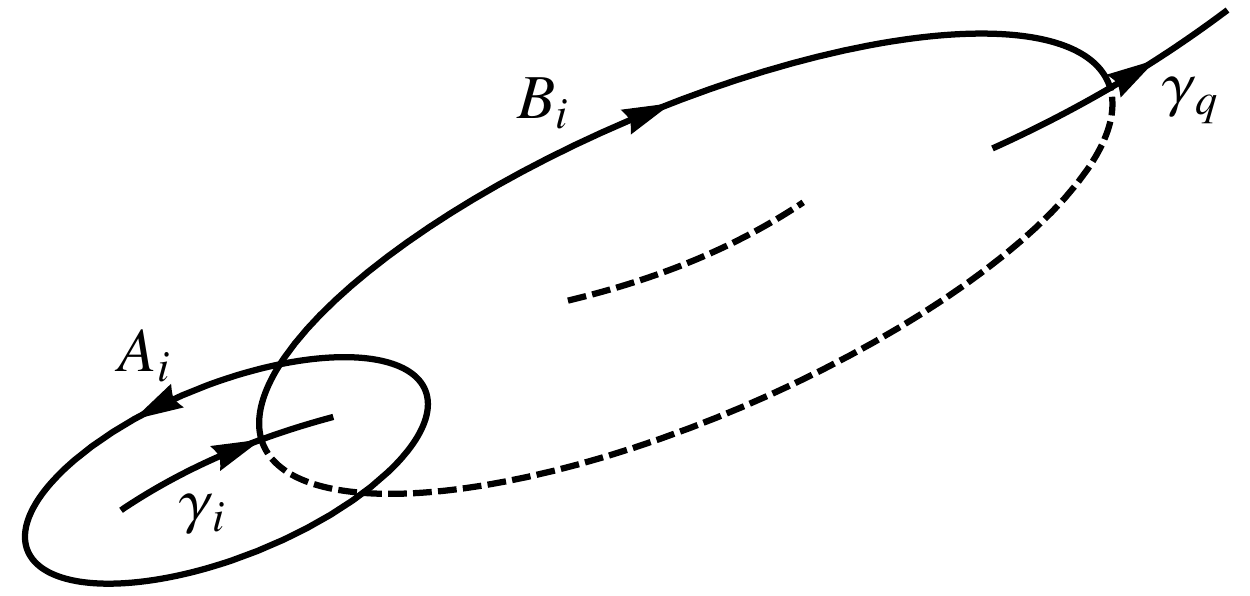}
    \end{center}
    \caption{The homology basis.
    \label{fig2}}
\end{figure}
We are going to decompose  $\rmd\mathbb{S}$ in terms of  the set Abelian differentials of first  kind (i.e., holomorphic)
$\{ \rmd \varphi_i\}_{i=1}^{q-1}$, second kind $\rmd \Omega_n$  $(n\geq 1)$ and third kind $\rmd \Omega_0$  described in
Appendix A. These Abelian differentials can be written as
\begin{equation}\label{dsc}
\rmd \varphi_j(z) = \frac{p_j(z)}{y_0(z)}\rmd z,\quad \rmd \Omega_n=
\left(\frac{n}{2} z^{n-1}+\frac{P_n(z)}{y_{0}(z)}\right)\rmd z,\quad \rmd\Omega_0 = \frac{P_0(z)}{y_{0}(z)} \rmd z,
\end{equation}
for appropriate polynomials $p_j(z)$ and $P_n(z)$.

We first observe that
\begin{equation}
  \label{1.4}
\mathrm{y}(z)\rmd z   = \left\{\begin{array}{ll}
                      \displaystyle\left(W'(z)-\frac{2s}{z}+\mathcal{O}(z^{-2})\right)\rmd z,
                      \quad\mbox{as $z\rightarrow \infty_1$},\\\\
                      \displaystyle\left(-W'(z)+\frac{2s}{z}+\mathcal{O}(z^{-2})\right)\rmd z,
                      \quad \mbox{as $z\rightarrow \infty_2$}.
                     \end{array}\right.
\end{equation}
Since the only poles of $\mathrm{y}(z)\rmd z$ are at $\infty_1$ and $\infty_2$, from~(\ref{1.4}) we deduce that
\begin{equation}
    \rmd\mathbb{S}-\sum_{n=1}^{N+1} g_n\, \rmd \Omega_n+s\,\rmd \Omega_0
\end{equation}
is a first-kind Abelian differential in $M$. Consequently it admits a decomposition in the canonical basis
\begin{equation}
    \label{dba}
    \rmd\mathbb{S}-\sum_{n=1}^{N+1} g_n\, \rmd \Omega_n+s\,\rmd \Omega_0
    =
    \sum_{j=1}^{q-1}\lambda_j \, \rmd \varphi_j,
\end{equation}
for some complex coefficients $\lambda_j\in \mathbb{C}$. Thus, we may write
\begin{equation}
    \label{idy}
    \mathrm{y}(z) \rmd z
    =
    -W'(z) \rmd z+2 \sum_{j=1}^{q-1} \lambda_j\, \rmd \varphi_j +2 \sum_{n=1}^{N+1} g_n\, \rmd \Omega_n-2s\,\rmd \Omega_0.
\end{equation}
Moreover, since $A_i(\rmd \Omega_n)=0$ for all $i=1,\ldots,q-1$ and $n\geq 0$ then from (\ref{toof}) and (\ref{idy}) we get
\[
\lambda_j=-2\pi\,\rmi\,s_j.
\]
Therefore we obtain  the decomposition
\begin{equation}\label{ds1}
\rmd\mathbb{S}=\sum_{n=1}^{N+1} g_n \rmd \Omega_n -\sum_{j=1}^{q}s_j \Big(\rmd \Omega_0+2\,\pi\,\rmi\, (1-\delta_{jq})\,\rmd \varphi_j\Big)
\end{equation}

The differential  $\rmd\mathbb{S}$ generates a system of Whitham equations with respect to the 't Hooft parameters ${\bf s}$ ~\cite{ch03}-~\cite{ch03b}.
To derive this system we notice that $\rmd\mathbb{S}$ has poles  at $\infty_1$ and $\infty_2$ only. Moreover
 its derivatives with respect to the 't~Hooft parameters satisfy
\[
\frac{\partial \rmd\mathbb{S}}{\partial s_j}= \left\{\begin{array}{ll}
                      \displaystyle\left(-\frac{1}{z}+\mathcal{O}(z^{-2})\right)\rmd z,
                      \quad\mbox{as $z\rightarrow \infty_1$},\\\\
                      \displaystyle\left(\frac{1}{z}+\mathcal{O}(z^{-2})\right)\rmd z,
                      \quad \mbox{as $z\rightarrow \infty_2$}.
                     \end{array}\right.
\]
and
\[
A_k\Big(\frac{\partial \rmd\mathbb{S}}{\partial s_j}\Big)=-2\,\pi\,\rmi\, \delta_{jk}, \quad k=1,\ldots,q-1,
\]
which imply
\begin{equation}\label{wh}
\frac{\partial \rmd\mathbb{S}}{\partial s_j}=-\rmd \Omega_0-2\,\pi\,\rmi\, (1-\delta_{jq})\,\rmd \varphi_j.
\end{equation}
As a consequence we have the following system of Whitham equations
\begin{equation}\label{ww}
\frac{\partial }{\partial s_i}\Big(\rmd \Omega_0+2\,\pi\,\rmi\, (1-\delta_{jq})\,\rmd \varphi_j\Big)=
\frac{\partial }{\partial s_j}\Big(\rmd \Omega_0+2\,\pi\,\rmi\, (1-\delta_{iq})\,\rmd \varphi_i\Big).
\end{equation}

Using (\ref{dsc})
we may  rewrite  (\ref{wh}) as
\begin{equation}\label{wh2}
\frac{\partial y(z)}{\partial s_j}=-2\,\frac{P_0(z)-2\,\pi\,\rmi\, (1-\delta_{jq})\,p_j(z)}{y_0(z)}.
\end{equation}
Hence  dividing  both members of the resulting equations by $y_0(z)$ and equating the residues at  $z=\beta_i$ we get the following
system of ordinary differential equations
\begin{equation}\label{wh3}
\frac{\partial \beta_i}{\partial s_j}=4\,\frac{P_0(\beta_i,\boldsymbol \beta)-2\pi\,\rmi\,
(1-\delta_{jq})\,p_j(\beta_i,\boldsymbol \beta)}{\prod_{l=1}^p(\beta_i-\alpha_l)\prod_{k\neq i}(\beta_i-\beta_k)},
\end{equation}
which will be henceforth referred to as the Whitham equations for the branch points.  From these equations it follows that for
spectral curves in the regular sectors the branch points are analytic functions of the 't Hooft parameters.

From (\ref{idy}) and using   (\ref{dsc}),  we have
\begin{equation}
    \label{eep}
    y(z) y_0(z) = 2 \sum_{n=1}^{N+1} g_n\, P_n(z) -2 s\,P_0(z)- 4\pi\, \rmi\sum_{j=1}^{q-1}s_j \,p_j(z).
\end{equation}
Therefore if we set $z=\beta_i$ in this identity we find the system
 \begin{equation}
  \label{ce1}
  \sum_{n=1}^{N+1} g_n\, P_n(\beta_i,\boldsymbol \beta) -  \sum_{j=1}^q s_j\Big(P_0(\beta_i,\boldsymbol \beta)+
  2\pi\, \rmi\,(1-\delta_{jq}) \, p_j(\beta_i,\boldsymbol \beta) \Big)=0 ,
\end{equation}
where $i=1,\ldots,2q$, which provides an implicit integration of the Whitham equations and determines the branch points in terms of
the 't Hooft parameters.   We observe that (\ref{ce1}) means that the generating differential $\rmd\mathbb{S}$
vanishes at the branch points $z=\beta_i$ ~\cite{KR04}.

 \subsection{Prepotentials and  $\tau$-functions }

 As we have said before, the present work deals with the moduli space of spectral curves  (\ref{sp0}) for a fixed polynomial
 $W(z)$, but  obviously,  the coefficients ${\bf g}=(g_1,\ldots,g_N)$ of $W(z)$ can be also  considered as free complex parameters
 too. In that case, repeating  the analysis of the last subsection we  immediately find
 \[
\frac{\partial \rmd\mathbb{S}}{\partial g_n}=\rmd \Omega_n,
\]
and consequently we  may enlarge the systems of Whitham equations (\ref{ww}) and (\ref{wh3}) including differential equations with
respect to the parameters ${\bf g}$.  In this sense it is important to notice the properties of the prepotential
functional as a function of these parameters. Indeed, from (\ref{free}) and (\ref{s1}),  taking into account that
$\rho(\beta_j)=0$, we have
\begin{eqnarray*}
        \frac{\partial \mathcal{F}}{\partial g_n}
    & = & \int_{\gamma} z^n \rho(z)| \rmd z|
    +\int_{\gamma}\Big(W(z)
    -
\int_{\gamma}\log(z-z')^2 \rho(z')| \rmd z'| \Big)\frac{\partial \rho(z)}{\partial g_n}| \rmd z|\\
        & = & \int_{\gamma}z^n \rho(z)| \rmd z|+
    \int_{\gamma} \left( W(z) - ( g(z_+)+g(z_-) ) \right) \frac{\partial  \rho(z)}{\partial g_n}| \rmd z| \\\nonumber
    & = &
     \int_{\gamma}z^n \rho(z)| \rmd z|+\sum_{j=1}^q l_j \int_{\gamma_j} \frac{\partial  \rho(z)}{\partial g_n} | \rmd z| \\
     & = &
     \int_{\gamma}z^n \rho(z)| \rmd z|+ \sum_{j=1}^q l_j\frac{\partial s_j}{\partial g_n} = \int_{\gamma}z^n \rho(z)| \rmd z|.
\end{eqnarray*}
On the other hand  from (\ref{y}) and  (\ref{ds}) it follows that as $z\rightarrow \infty_1$
\[
\rmd\mathbb{S}=\Big(W'(z) -\sum_{n\geq 1}\frac{1}{z^n}\int_{\gamma}(z')^{n-1} \rho(z')|  \rmd z' |\Big) \rmd z.
\]
Then we conclude that
\begin{equation}\label{tau}
\frac{\partial \mathcal{F}}{\partial g_n}=-\frac{1}{2\pi\,\rmi}\oint_{\Gamma_{\infty}}z^n\rmd\mathbb{S},
\end{equation}
where $\Gamma_{\infty} $ is a positively oriented circle of large radius.

 From ~(\ref{s1}) and~(\ref{two}) we find that the parameters $l_i$  are related to the $B$-periods in the form
\begin{equation}
    \label{perb}
    \oint_{B_i}\rmd\mathbb{S}  = l_q-l_i ,\quad i=1,\ldots,q-1,
\end{equation}
so that using (\ref{ele}) we have
\begin{equation}
\label{perb}
\frac{\partial\mathcal{F}}{\partial s_q}-\frac{\partial\mathcal{F}}{\partial s_i}=\oint_{B_i} \rmd\mathbb{S},\quad i=1,\ldots,q-1,
\end{equation}
 The relations (\ref{tau}) and (\ref{perb})  mean  that the logarithm $\log\mathcal{F}$ of the prepotential functional
 is a $\tau$-function of the Whitham hierarchy with respect to the variables $({\bf g},{\bf s})$.

\section{Spectral curves in terms of deformation parameters}

In the previous section  the set  ${\bf s}$ of  't Hooft parameters has been used to describe the integrable structure which underlies spectral
curves.
However, the use of the alternative set ${\bf t}$ of deformation parameters turns to be more convenient  for dealing with several problems as,
for example,  the construction of explicit examples of multicut spectral curves.

\subsection{Roots of spectral curves as functions of deformation parameters}
The  spectral curves in  $\mathcal{M}_q$ satisfy  the system of $2N$ equations
\begin{equation}\label{roots}
\left\{\begin{array}{cc}P(\alpha_l,{\bf t})= P'(\alpha_l,{\bf t})=0,&
l=1,\ldots,p,\\Ê\\ P(\beta_i,{\bf t})=0, & i=1,\ldots,2q,\end{array}\right.
\end{equation}
where
 $P(z,{\bf t})$ denotes the polynomial
\begin{equation}\label{polp}
P(z,{\bf t})=y(z,{\bf t})^2=W'(z)^2+\sum_{k=1}^{N}t_k\,z^{N-k}.
\end{equation}
This system defines an affine variety in the space $\mathbb{C}^{2N+q}$ of the variables $(\alpha_l,\beta_i,t_k)$.
The Jacobian of the system  (\ref{roots})
\begin{equation}\label{jac1}
\frac{\partial\Big(P(\alpha_1),\cdots ,P(\alpha_p),P'(\alpha_1),\cdots ,P'(\alpha_p),P(\beta_1),\cdots ,P(\beta_{2q})  \Big)}
{\partial\Big(\alpha_1,\cdots ,\alpha_p, \beta_1,\cdots,\beta_{2q}, t_1,\cdots ,t_N \Big)}
\end{equation}
evaluated at solutions of (\ref{roots}) is of the form
\begin{equation}
\left(\begin{array}{c|c}A&B\\ \hline C&D
\end{array}\right),
\end{equation}
where $A$ is the $p\times ( p+2 q)$  zero  matrix  since from (\ref{roots})
\[
A_{lm}=\delta_{lm} P'(\alpha_l)= 0,\quad  l=1,\ldots,p,\;\;m=1,\ldots,p+2q.
\]
Furthermore $B$ is the Vandermonde matrix
\[
B=\left(\begin{array}{cccc}\alpha_1^{N-1} & \alpha_1^{N-2} & \cdots & 1 \\\alpha_2^{N-1} & \alpha_2^{N-2} &
\cdots & 1 \\\vdots& \vdots& \ddots & \vdots \\ \alpha_p^{N-1} & \alpha_p^{N-2} & \cdots & 1\end{array}\right),
\]
$C$ is the diagonal matrix
\[
C={\rm diag}\Big(P''(\alpha_1),\cdots, P''(\alpha_p),\;P'(\beta_1),\cdots ,P'(\beta_{2q})\Big),
\]
and
\[
D=\left(\begin{array}{cccc}(N-1)\alpha_1^{N-2} & (N-2)\alpha_1^{N-3} & \cdots & 0\\
(N-1)\alpha_2^{N-1} & (N-2)\alpha_2^{N-3} & \cdots & 0 \\\vdots& \vdots& \ddots & \vdots \\ (N-1)\alpha_p^{N-2} & (N-2)\alpha_p^{N-3} & \cdots & 0\\
\beta_1^{N-1} & \beta_1^{N-2} & \cdots & 1 \\\beta_2^{N-1} & \beta_2^{N-2} & \cdots & 1 \\\vdots& \vdots& \ddots & \vdots \\ \beta_{2q}^{N-1} &
\beta_{2q}^{N-2} & \cdots & 1
\end{array}\right)
\]
Let us consider now spectral curves in the regular sector  ${\rm reg}\mathcal{M}_q$. It is easy to see that for any set of  $q$ deformation
parameters $(t_{i_1},\cdots,t_{i_q})$ the minor of the Jacobian corresponding to the derivatives with respect to the variables $(\alpha_l,\beta_i)$
and the remaining $N-q$ deformation parameters $t_k\,(k\notin\{i_1,\ldots,i_q\})$ is a non singular  $(2N)\times (2N)$ matrix. Therefore the
implicit function Theorem implies
\begin{theo} For spectral curves in the regular sector  ${\rm reg}\mathcal{M}_q$,  each set of $q$ different  deformation parameters
$(t_{i_1},\cdots,t_{i_q})$ defines an analytic system of coordinates, i.e.  all the variables $(\alpha_l,\beta_i,t_k)$ are local analytic
functions of $(t_{i_1},\cdots,t_{i_q})$

\end{theo}

  If we eliminate the variables $(\alpha_l,\beta_i)$ in the system (\ref{roots}) or, alternatively, in the system obtained identifying
  coefficients of powers of $z$ between (\ref{sp0}) and (\ref{phys}),  we get a set of
 $N-q$ constraints
 \begin{equation}\label{man}
f_k(\bt )=0,\quad k=1,\ldots,N-q,
\end{equation}
which characterizes the class $\mathcal{M}_q$
 in the space  $\mathbb{C}^N$ of  deformation parameters.
These constraints can be used to define analytic systems of coordinates  in ${\rm reg}\mathcal{M}_q$ given  by  sets
$(t_{i_1},\cdots,t_{i_q})$ of $q$ distinct deformation parameters.

 The class $\mathcal{M}_N$ contains all  the  spectral curves (\ref{sp0}) and  will be called \emph{the generic class}.
 Its  elements can be factorized in the generic form
\begin{equation}\label{facg}
y(z,\bt)^2=\prod_{j=1}^{2N}(z-\beta_j),
\end{equation}
where the roots $\beta_i$ may coincide.
In the case of  the generic class  $\mathcal{M}_N$  the set of the deformation parameters  $(t_1,\ldots,t_N)$ defines
an analytic system of coordinates  in the regular sector ${\rm reg}\mathcal{M}_N$.
Obviously the domain in the space of  deformation parameters $\mathbb{C}^N$ which corresponds to  ${\rm reg}\mathcal{M}_N$
is the open set determined  by
\begin{equation}\label{discg}
D(\bt)\neq 0,
\end{equation}
where $D=D(\bt)$ is the discriminant of the polynomial $y(z,\bt)^2$. Consequently ${\rm sing}\mathcal{M}_N$ is represented
by the hypersurface  determined by the constraint
\begin{equation}\label{discg2}
D(\bt)= 0.
\end{equation}

From the point of view of the applications to gauge/string dualities it is important  to determine under what conditions a map
$(t_{i_1},\cdots,t_{i_q})\mapsto (s_1,\ldots,s_q) $ relating the coordinate charts of deformation parameters and 't Hooft
parameters defines an  analytic change of variables. Now   from (\ref{toof})  we conclude that the  Jacobian matrix of these
transformation  is given by
\[
\frac{\partial s_j}{\partial t_{i_k}}=-\frac{1}{8\,\pi\,\rmi}\oint_{A_j}\Big(z^{N-i_k}+\sum_{i\notin \{i_1,\ldots,i_q\}}z^{N-i}\frac{\partial t_i}
{\partial t_{i_k}}\Big)\frac{{\rm d}z}{y(z)}.
\]
Thus the question reduces to the verification of the nonsingularity of  this Jacobian matrix.
For spectral curves in ${\rm reg}\mathcal{M}_N$
it follows   from
(\ref{toof})  that
\[
\frac{\partial s_j}{\partial t_k}=-\frac{1}{8\,\pi\,\rmi}\oint_{A_j}\frac{z^{N-k}}{\sqrt{\prod_{j=1}^{2N}(z-\beta_j)}}\,{\rm d}z,\quad j,k=2\,\ldots,N,
\]
This means that the Jacobian of the transformation
$(t_2,\ldots,t_N)\mapsto(s_2,\ldots,s_N)$ is non singular since it is the matrix of $A$-periods of a basis of Abelian differentials in
the Riemann surface  given by
\[
y(z)^2=\prod_{j=1}^{2N}(z-\beta_j),
\]
where all the branch points are different ($N-1$ genus).   Moreover,  according to (\ref{asb}) it follows at once that $t_1=-4s$ , then  the change of variables $(t_1,\ldots,t_N)\mapsto(s_1,\ldots,s_N)$ is an  analytic  diffeomorphism locally around any point in the
regular sector ${\rm reg}\mathcal{M}_q$.  Consequently using Theorem 1 we conclude that the set of 't~Hooft parameters $(s_1,\ldots,s_N)$
provides an analytic system of coordinates in  ${\rm reg}\mathcal{M}_N$.

\subsection{Whitham equations with respect to deformation parameters}
We have seen that for spectral curves in the  regular sector  ${\rm reg}\mathcal{M}_q$ all the variables  $(\alpha_l,\beta_i,t_k)$ can be
expressed as analytic functions in any coordinate system  of $q$   deformation parameters. In particular one may consider the expressions
for the simple roots in terms of the coordinates $(t_1,\ldots,t_q)$
\begin{equation}\label{diffb}
\beta_i=\beta_i(t_1,t_2,\ldots, t_q).
\end{equation}
We are going to derive the system of Whitham equations which characterizes the functions (\ref{diffb}). Indeed, taking into account  that
the double roots $\alpha_l$ are distinct then from (\ref{polp}) we get the identities
\begin{equation}\label{idt}
t_{l+q}=\sum_{m=1}^p V_{lm}(\boldsymbol \alpha)\,\Big[ P(\alpha_m,\bt)-\Big(W'(\alpha_m)^2+\sum_{j=1}^q t_j\,\alpha_m^{N-j} \Big)\Big],
\end{equation}
for $ l=1,\ldots,p=N-q$ where $V_{lm}(\boldsymbol \alpha)$ is the inverse of the Vandermonde matrix
$
(\alpha_{l}^{p-m})_{l,m=1}^p
$.
As a consequence the functions
\begin{equation}\label{redp}
P(\beta_i,{\bf t})-\sum_{l,m=1}^p \beta_i^{p-l}\, V_{lm}(\boldsymbol \alpha)\,P(\alpha_{m},{\bf t})
\end{equation}
do not depend on the deformation parameters $(t_{q+1},\ldots,t_ N)$ since they can be written as
\[
Q(\boldsymbol \alpha,\beta_i)+\sum_{j=1}^q t_j\, \Big(\beta_i^{N-j}- \sum_{l,m=1}^p \beta_i^{p-l}\,
V_{lm}(\boldsymbol \alpha)\,\alpha_{m}^{N-j}  \Big),
\]
where
\[
Q(\boldsymbol \alpha,\beta)=W'(\beta)^2-\sum_{l,m=1}^p \beta^{p-l}\, V_{lm}(\boldsymbol \alpha)\,W'(\alpha_m)^2.
\]
Now for  spectral  curves in ${\rm reg}\mathcal{M}_q$ we may express all the variables $(\alpha_l,\beta_i,t_k)$ as functions of the coordinates
$(t_1,\ldots,t_q)$. Hence
differentiating the equations
\[
P(\beta_i,{\bf t})-\sum_{l,m} \beta_i^{p-l}\, V_{lm}(\boldsymbol \alpha)\,P(\alpha_{m},{\bf t})=0
\]
 with respect to the free parameters $\{t_j\}_{j=1}^q$, taking into account that $P(\alpha_l,{\bf t})=P'(\alpha_l,{\bf t})=0$,  we obtain
 the system of differential equations
\begin{equation}\label{diff}
\frac{\partial \beta_i}{\partial t_j}=\frac{\sum_{l,m} \beta_i^{p-l}\, V_{lm}(\boldsymbol \alpha)\,\alpha_{m}^{N-j}-\beta_i^{N-j}}{\prod_l
(\alpha_l-\beta_i)^2\prod_{j\neq i}(\beta_i-\beta_j)}.
\end{equation}
 This is the system of Whitham equations for the branch points with respect to the deformation parameters.

The system (\ref{diff}) shows how
the functions $\beta_i(t_1,\ldots,t_q)$ become singular for spectral curves in the singular sector ${\rm sing}\mathcal{M}_q$, where at least
two roots of the set $(\alpha_l,\beta_i)$ coalesce.

\vspace{0.3cm}

For  $q=N$ and $q=N-1$ the system reduces to
\begin{equation}\label{sn}
\frac{\partial \beta_i}{\partial t_j}=-\frac{\beta_i^{N-j}}{\prod_{j\neq i}(\beta_i-\beta_j)},
\end{equation}
and
\begin{equation}\label{snm}
\frac{\partial \beta_i}{\partial t_j}=\frac{\alpha^{N-j}-\beta_i^{N-j}}{(\alpha-\beta_i)^2\prod_{j\neq i}(\beta_i-\beta_j)},
\quad \Big(\alpha=-Ng_N-\frac{\beta_1+\cdots+\beta_{2q}}{2}\Big)
\end{equation}
respectively.

\subsection{The  Gaussian and cubic models }

A complete explicit analysis of the classes $\mathcal{M}_q$ is possible only for  polynomials $W(z)$ of low degree. We next discuss some
examples for the cases of degrees two and three.  Similar studies  have been recently performed  for the cubic model  in ~\cite{MAR10} and for the quartic model in ~\cite{BE12} .

\paragraph{Gaussian model}

For the Gaussian model
\begin{equation}
    \label{gau}
    W(z) = \frac{z^2}{2},
\end{equation}
we have $N=1$ and $y(z,{\bf t})^2=z^2-4s$.  The system (\ref{roots}) for $q=1$ and $q=0$ reduce to
\[
\beta_i^2-4s=0,\quad i=1,2,
\]
and
\[
\alpha^2-4s=0,\quad \alpha=0,
\]
respectively.
Thus for $\mathcal{M}_1$ the branchpoints are
\begin{equation}
    \beta_1 = -\beta_2=2\,\sqrt{s}.
\end{equation}
The regular sector   ${\rm reg}\mathcal{M}_1$  is given by  the spectral curves corresponding to values  $s\neq 0$. The singular sector
${\rm sing}\mathcal{M}_1$ coincides with  $\mathcal{M}_0$ and contains only the spectral curve $y(z)^2=z^2$ which corresponds to $s=0$.

\paragraph{Cubic model}
Let us consider the cubic model
\begin{equation}
    \label{eq:w3}
    W(z) = \frac{z^3}{3} - g\, z,\quad g\in \mathbb{C}\setminus \{0\}.
\end{equation}
 In this case $N=2$ and
\begin{equation}
    \label{y3}
    y(z,{\bf t})^2 =(z^2-g)^2+t_1\,z+t_2,\quad t_1=-4 s.
\end{equation}
The generic  class $\mathcal{M}_2$  is the set of all the spectral curves (\ref{y3}). If we write
\begin{equation}\label{y32}
y(z,{\bf t})^2=(z-\beta_1)(z-\beta_2)(z-\beta_3)(z-\beta_4),
\end{equation}
introduce
the variables
\begin{eqnarray}\label{nv}
\nonumber u_1=\frac{\beta_1+\beta_2}{2},\quad u_2=\frac{\beta_3+\beta_4}{2},\\Ê\\
\nonumber \delta_1=\frac{\beta_1-\beta_2}{2},\quad \delta_2=\frac{\beta_3-\beta_4}{2}.
\end{eqnarray}
and identify the coefficients of powers of $z$  in equations (\ref{y3}) and (\ref{y32}) we obtain the relations
\begin{equation}\label{sys22}
 u_2=-u_1,\quad \delta_1^2=g-u_1^2+\frac{s}{u_1},\quad  \delta_2^2= g-u_1^2-\frac{s}{u_1},
\end{equation}
and the equation which determines $u_1$ as a function of the deformation parameters $(s,t_2)$
\begin{equation}\label{u1}
4u_1^6-4 g u_1^4-t_2 u_1^2-s^2=0.
\end{equation}
The regular sector ${\rm reg }\mathcal{M}_2$ is
 determined by  the points $(s,t_2)\in \mathbb{C}^2$ at which the discriminant of $y(z,{\bf t})^2$  does not vanish, i.e.
\begin{equation}\label{discn}
{\rm reg }\mathcal{M}_2 \leftrightarrow \{(s,t_2)\in \mathbb{C}^2\,:\, 27 s^4+(18 g t_2+16 g^3)\,s^2-t_2^3-g^2 t_2^2 \neq 0\}.
\end{equation}
The singular sector ${\rm sing }\mathcal{M}_2$  is given by the spectral curves  (\ref{y3}) which have a multiple root. Their  possible
sets of root multiplicities are
\[
 [2_1,1_2],\, [1_1,1_3],\,[2_2],\,[1_4],
\]
which obviously means that
\[
{\rm sing }\mathcal{M}_2=\mathcal{M}_1.
\]

Let us now consider  the  class  $\mathcal{M}_1$
\begin{equation}
    \label{eq:w3y1}
    y(z,\bt)^2= (z-\alpha)^2 (z-\beta_1)(z-\beta_2).
\end{equation}
Equating (\ref{y3}) and (\ref{eq:w3y1}),  identifying coefficients  and  using the variables
\[
u=\frac{\beta_1+\beta_2}{2},\quad \delta=\frac{\beta_1-\beta_2}{2},
\]
we get the system of equations
\begin{equation} \label{sys1}
\left\{\begin{array}{c}\alpha=-u,\quad 2 u^2 + \delta^2 = 2 g, \\u\, \delta^2 = 2 s, \quad u^4-u^2\,\delta^2-g^2=t_2.\end{array}\right.
\end{equation}
Therefore $u$  satisfies the cubic equation
\begin{equation}
    \label{eq:beta}
    u^3 - g u + s = 0,
\end{equation}
and $\delta^2$ is determined by
\begin{eqnarray}
    \label{eq:delta1}
    \delta^2 & = & \frac{2 s}{u}, \quad \mbox{if $u\neq0$},\\
    \label{eq:delta2}
    \delta^2 & = & 2 g, \quad \mbox{if $u=0$}.
\end{eqnarray}
Moreover, eliminating $u$ and $\delta^2$ in (\ref{sys1}) we obtain a  constraint for the deformation parameters $(s,t_2)$ which, as it should be
expected, is equivalent to the vanishing of the discriminant  of $y(z,{\bf t})^2$
\begin{equation}\label{disc}
\mathcal{M}_1 \leftrightarrow \{(s,t_2)\in \mathbb{C}^2\,:\, 27 s^4+(18 g t_2+16 g^3)\,s^2-t_2^3-g^2 t_2^2 = 0\}.
\end{equation}

According to (\ref{sys1}) $\beta_1=\beta_2$ if and only if  $(s,t_2)=(0,0)$, i.e. $y^2=(z^2-g)^2$. Moreover if $\alpha$  coincides with $\beta_1$
or $\beta_2$  then the polynomial $y(z)^2$ acquires a  root of multiplicity $m\geq 3$ and therefore  it satisfies ~\cite{GI82}
\[
\sum_{k=0}^4 (-1)^k D_z^k y^2(z)\, D_z^{4-k} y^2(z)=0,
\]
which gives the constraint
\begin{equation}\label{giv}
3\,t_2+4\,g^2=0,
\end{equation}
or, due to (\ref{disc}) equivalently,
\begin{equation}\label{giv2}
(s,t_2)=\Big(\pm 2\,\Big(\frac{g}{3}\Big)^{3/2},-\frac{4}{3}\, g^2\Big).
\end{equation}
 Therefore, the singular sector ${\rm sing}\mathcal{M}_1$ is given by only three points of the space of deformation parameters
\begin{equation}\label{giv3}
{\rm sing}\mathcal{M}_1 \leftrightarrow\{(s,t_2)=(0,0),\,\Big(\pm 2\,\Big(\frac{g}{3}\Big)^{3/2},-\frac{4}{3}\, g^2\Big)\}.
\end{equation}
 The corresponding root multiplicities are $[2_2]$ for $(0,0)$ and $[1_1,1_3]$ for $(\pm 2\,(g/3)^{3/2},-4 g^2/3)$. Figure 4 exhibits an example
 of a  curve in  ${\rm sing}\mathcal{M}_1$ in which a triple root is formed as the outcome of a merging of a simple root and the double root.
\begin{figure}
    \begin{center}
        \includegraphics[width=8cm]{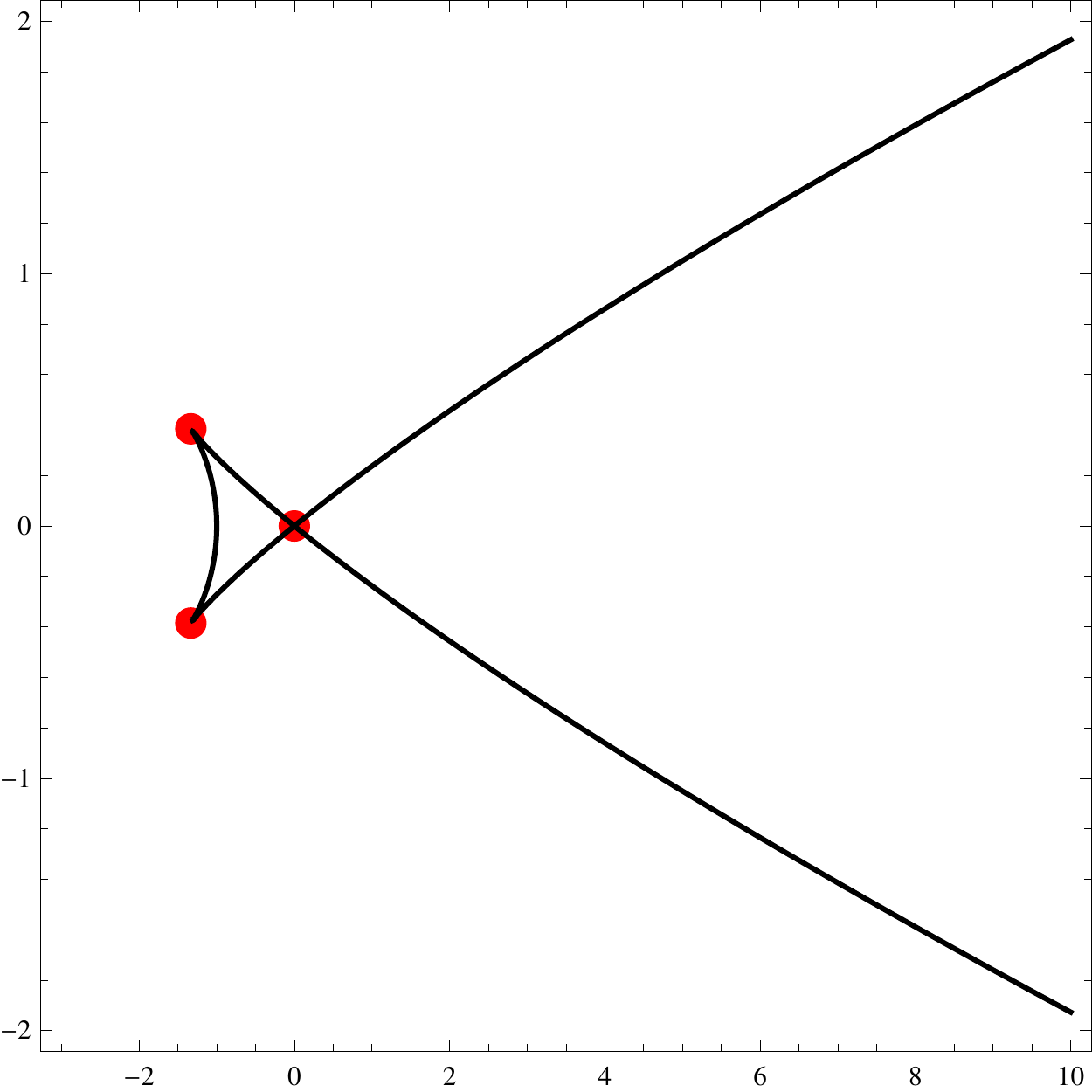}
    \end{center}
    \caption{The classes of spectral curves of the cubic model for $g=1$ in the space of real values of $(t_2,s)$. The whole real plane describes
    $\mathcal{M}_2$ ,  the curve corresponds to the implicit equation (\ref{disc}) and represents  ${\rm sing}\mathcal{M}_2=\mathcal{M}_1$, the
    three marked points represent ${\rm sing}\mathcal{M}_1$. The point $(0,0)$  represents  $\mathcal{M}_0$.
    \label{fig3}}
\end{figure}

From (\ref{eq:beta}) it follows that
the  class $\mathcal{M}_1$ decomposes into three subsets
\[
\mathcal{M}_1=\cup_{k=0}^2 \mathcal{M}_1^{(k)}
\]
which are determined
 by  the three branches
\begin{equation}
    \label{eq:betak}
    u_k(s) = - \frac{g}{3 \Delta_k(s)} - \Delta_k(s), \quad (k=0,1,2)
\end{equation}
where
\begin{equation}
    \Delta_k(s) = \rme^{\rmi 2\pi k/3}
                         \sqrt[3]{\frac{s}{2} + \sqrt{\frac{s^2}{4} - \left(\frac{g}{3}\right)^3}}.
\end{equation}
The
expansions of $u_k(s)$ as $s\rightarrow 0$ are
\begin{eqnarray}
    \label{eq:b0s}
    u_0(s) & = & - \sqrt{g} - \frac{s}{2g} + \frac{3 s^2}{8 g^{5/2}} + \mathcal{O}(s^3),\\
    \label{eq:b1s}
    u_1(s) & = &   \sqrt{g} - \frac{s}{2g} - \frac{3 s^2}{8 g^{5/2}} + \mathcal{O}(s^3),\\
    \label{eq:b2q}
    u_2(s) & = & \frac{s}{g}+\mathcal{O}(s^3),
\end{eqnarray}
which show that in this limit the branches $u_0$ and $u_1$ represent families of one-cut spectral curves in which
the cuts shrink to the corresponding critical point $-\sqrt{g}$ and $+\sqrt{g}$ of $W(z)$.

 The solution $u_2(s)$ has a different behavior
as $s\rightarrow 0$, because in this limit $u_2\rightarrow 0$, $\delta^2\rightarrow 2 g$ and  therefore the branch  points $\beta_1$ and $\beta_2$ reduce to two different points  $\sqrt{2 g}$ and $-\sqrt{2g}$. Note also from (\ref{sys1})  that $t_2\rightarrow -g^2$ as $s\rightarrow 0$.
\begin{figure}
    \begin{center}
        \includegraphics[width=10cm]{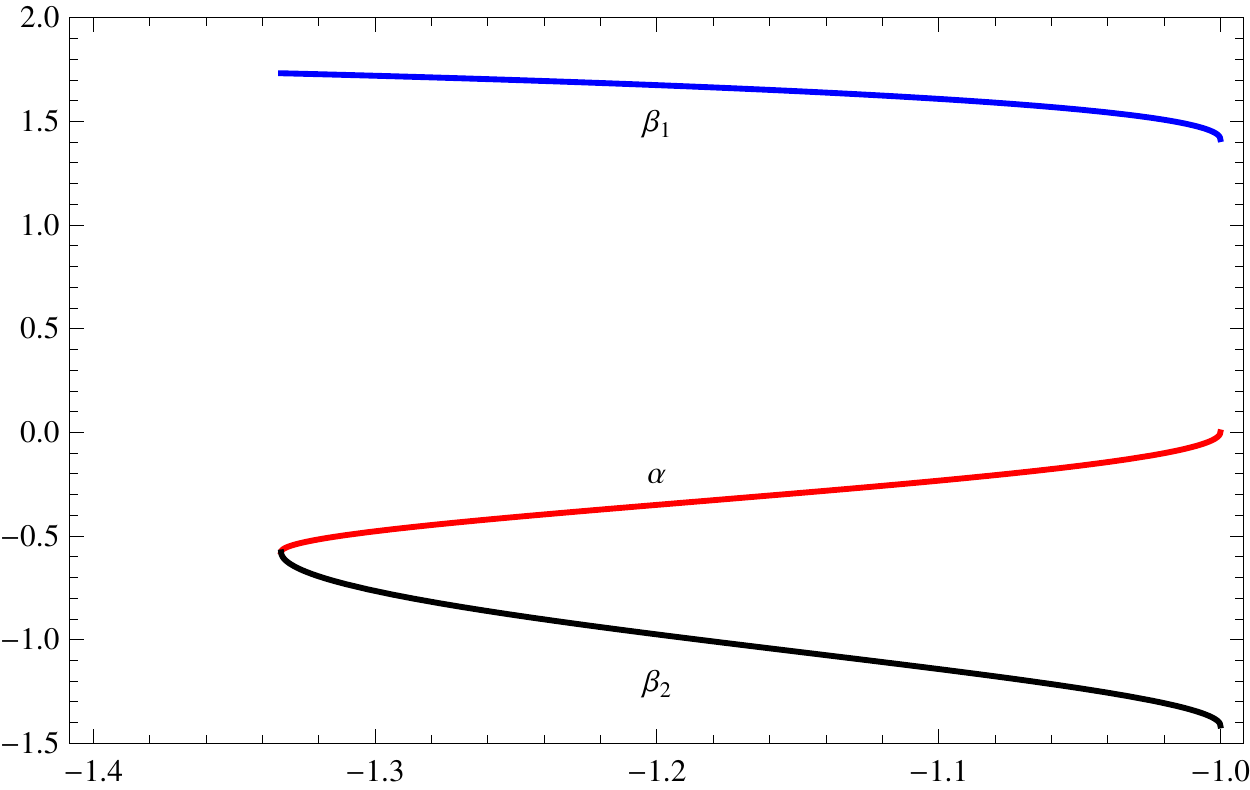}
    \end{center}
    \caption{The functions $\alpha(t_2)$ and $\beta_i(t_2)$
at a merging of the double root and a simple root in the cubic model for $g=1$.
    \label{fig4}}
\end{figure}

Finally, the class $\mathcal{M}_0$  contains only one   spectral curve  $y^2=(z^2-g)^2$ which  corresponds to $(s,t_2)=(0,0)$.

\section{The Euler-Poisson-Darboux equation and the reduced  systems }

As we have seen the regular sectors  ${\rm reg}\mathcal{M}_q$ are completely characterized by the functions (\ref{diffb}). We will now prove
that these functions  can be determined  from a system of $2q$ equations for the variables
\[
 \boldsymbol \beta=(\beta_1,\ldots,\beta_{2q}),\quad  \bt^{(q)}=(t_1,\ldots,t_q)
\]
  These systems will be called \emph{reduced  systems}.  They are determined from a particular polynomial solution of the Euler-Poisson-Darboux
  equation
$E(-\frac{1}{2},-\frac{1}{2})$ (see
e.g. ~\cite{DAR})
\begin{equation}
2\,(\beta_{j}-\beta_{i})\,\frac{\partial ^{2}\mathbb{W}}{\partial \beta_{i}\,\partial
\beta_{j}}=\frac{\partial \mathbb{W}}{\partial \beta_{i}}-\frac{\partial \mathbb{W}}{
\partial \beta_{j}}.  \label{eulp}
\end{equation}

\subsection{Reduced systems}

Let us define the function
\begin{equation}
\mathbb{W}(\boldsymbol \beta,\bt)=\oint_{\Gamma_{\infty} }
y_{0}(z,\boldsymbol \beta)\,y(z,\bt)\,\frac{\rmd z}{2\,\pi \,\rmi},  \label{susper0}
\end{equation}
where
\[
y_{0}(z,\boldsymbol \beta)=\sqrt{\prod_{i=1}^{2q}(z-\beta_{i})},\quad y(z,\mathbf{t})=
\sqrt{W'(z)^2+f(z,\mathbf{t})}.
\]
and $\Gamma_{\infty} $ is a positively oriented circle of large radius. It obviously
satisfies the Euler-Poisson-Darboux equation (\ref{eulp}). Moreover, using
the expansion of $y(z,\mathbf{t})$ at $z\rightarrow \infty $, i.e.
\[
y(z,\mathbf{t})=\sqrt{W'(z)^2+f(z,\mathbf{t})}=W^{\prime }(z)+\frac{1}{2
}\frac{f(z,\mathbf{t})}{W^{\prime }(z)}+\mathcal{O}\Big(\frac{1}{z^{N+2}}
\Big),
\]
we deduce that $\mathbb{W}$ reduces to
\begin{eqnarray} \label{susper2}
\nonumber\mathbb{W}(\boldsymbol \beta,\bt)=\oint_{\Gamma_{\infty} }
\,W^{\prime }(z)\,y_{0}(z,\boldsymbol \beta)\,\frac{\rmd z}{2\,\pi \,\rmi}\\\\
\nonumber +\frac{1}{2}\,\sum_{j=1}^{q}t_{j}\,
\oint_{\Gamma_{\infty} }z^{N-j}\frac{y_{0}(z,
\boldsymbol \beta)}{W^{\prime }(z)}\,\frac{\rmd z}{2\,\pi \,\rmi}+\frac{t_{q+1}}{2}.
\end{eqnarray}
Observe  that  $\mathbb{W}-t_{q+1}/2$ does not depend
on $(t_{q+1},\ldots ,t_N)$ and that it
is  linear in $(t_1,\ldots ,t_q)$.
\vspace{0.3cm}
\subsubsection*{Examples:} In the  cubic model we have for $q=1$
\begin{equation*}
\mathbb{W}=\frac{1}{128} \Big[-\left(\beta _1-\beta
   _2\right){}^2 \left(5 \beta _1^2+6
   \beta _2 \beta _1+5 \beta _2^2-16
   g\right)-32 \left(\beta _1+\beta
   _2\right) t_1+64 t_2\Big]
\end{equation*}
and for $q=2$
\begin{eqnarray*}
\mathbb{W}=\frac{1}{256} \Big[2  \left(\beta _1+\beta _2+\beta
   _3+\beta _4\right) \left(\beta _1-\beta
   _2\right)^2\left(\beta _3-\beta
   _4\right)^2\\
   +\left(\beta _1-\beta_2\right){}^2\Big(-7 \beta _1^3-9 \beta _2
   \beta _1^2+5 \beta _3 \beta _1^2+5 \beta _4
   \beta _1^2-9 \beta _2^2 \beta _1\\
   +6 \beta _2
   \beta _3 \beta _1+6 \beta _2 \beta _4 \beta
   _1
   -7 \beta _2^3+5 \beta _2^2 \beta _3+5 \beta
   _2^2 \beta _4\Big)\\
   +\left(\beta _3-\beta
   _4\right){}^2 \Big(-7 \beta _3^3+5 \beta _1
   \beta _3^2+5 \beta _2 \beta _3^2-9 \beta _4
   \beta _3^2-9 \beta _4^2 \beta _3\\
   +6 \beta _1   \beta _4 \beta _3+6 \beta _2 \beta _4 \beta
   _3-7 \beta _4^3+5 \beta _1 \beta _4^2+5 \beta
   _2 \beta _4^2\Big)\Big]\\
   +\frac{1}{16}
   \left(\beta _1+\beta _2-\beta _3-\beta
   _4\right) \left(\beta _1-\beta _2+\beta
   _3-\beta _4\right) \left(\beta _1-\beta
   _2-\beta _3+\beta _4\right) g\\
   +\frac{1}{16}
   \Big[-\left(\beta _1-\beta _2+\beta
   _3-\beta _4\right){}^2+4 \left(\beta _1 \beta
   _3+\beta _2 \beta _4\right)+8
   g\Big]\, t_1\\
   -\frac{1}{4} \left(\beta _1+\beta_2+\beta _3+\beta _4\right)\, t_2
\end{eqnarray*}

We are now ready to prove the following characterization of the functions (\ref{diffb})
\begin{theo}
For spectral curves in  $\mathcal{M}_q$  the functions $\boldsymbol \beta= \boldsymbol \beta(\emph{\bt}^{(q)})$  satisfy
\begin{equation}
\frac{\partial \mathbb{W}}{\partial \beta_{i}}=0 \,\, \mbox{at  $\boldsymbol \beta=\boldsymbol \beta(\emph{\bt}^{(q)})$},\quad  i=1,\ldots ,2 q.
\label{criti0}
\end{equation}
\end{theo}

\noindent \textbf{Proof} One has
\[
\frac{\partial y_{0}(z,\boldsymbol \beta)}{\partial \beta_{i}}=-\frac{1}{2}\frac{
y_{0}(z,\boldsymbol \beta)}{z-\beta_{i}},
\]
so
\[
\frac{\partial \mathbb{W}}{\partial \beta_{i}}=-\frac{1}{2}\,\oint_{\Gamma_{\infty} }
\frac{y_{0}(z,\boldsymbol \beta)\,y(z,\bt)}{
z-\beta_{i}}\,\frac{\rmd z}{2\,\pi \,\rmi}.
\]
Moreover  for spectral curves in $\mathcal{M}_q$ we have that
\begin{equation}\label{yw}
\frac{y_{0}(z,\boldsymbol\beta )\,y(z,\bt)}{z-\beta_i}=\prod_{l=1}^p(z-\alpha
_{l})\,\prod_{j\neq i}(z-\beta _{j}),
\end{equation}
and then the statement follows.

 The equations (\ref{criti0}) can be expressed as
\[
\oint_{\Gamma_{\infty}}\frac{\prod_{j\neq i}(z-\beta_j)}{y_{0}(z,\boldsymbol \beta)}
\Big(W'(z)+\frac{1}{2}\frac{\sum_{k=1}^qt_k z^{N-k}}{W'(z)} \Big)\,\frac{\rmd z}{2\,\pi \,\rmi}=0,\quad i=1,\ldots 2q,
\]
and therefore for distinct $\beta_i $ they are equivalent to the system
\begin{equation}\everymath{\displaystyle}\label{criti2}
\left\{\begin{array}{l} \oint_{\Gamma_{\infty}}\frac{z^k}{\sqrt{\prod_{j=1}^{2q}(z-\beta_j)}}\Big(W'(z)+
\frac{1}{2}\frac{\sum_{j=1}^qt_j z^{N-j}}{W'(z)} \Big)\,\frac{\rmd z}{2\,\pi \,\rmi}=0,\\ \\ k=0,\ldots 2q-1\end{array}\right.
\end{equation}
We will refer to (\ref{criti0})  as the \emph{reduced system} for the $3q$ variables $(\boldsymbol \beta,\bt^{(q)})$. It must be observed
that  the  previous system  (\ref{roots})   for $(\alpha_l,\beta_i,t_k)$   involves a larger number $2p+3q$ of  variables .

\subsubsection*{Examples:}

For $q=1$  the reduced system (\ref{criti0}) is  the well-known  pair of equations which determines the planar limit of random matrix models
~\cite{DI95}
\begin{equation}\everymath{\displaystyle}\label{inte11}
\left\{\begin{array}{c} \oint_{\Gamma_{\infty}}\frac{W'(z)}{\sqrt{(z-\beta_1)(z-\beta_2)}} \rmd z=0,\\ \\
\oint_{\Gamma_{\infty}}z\,\frac{W'(z)}{\sqrt{(z-\beta_1)(z-\beta_2)}} \rmd z=4\,\pi\,\rmi\,s. \end{array}\right.
\end{equation}

For  $q=2$
the reduced system (\ref{criti0}) takes the form
\begin{equation}\everymath{\displaystyle}\label{inte112}
\left\{\begin{array}{l}
\oint_{\Gamma_{\infty}}\frac{W'(z)}{\sqrt{\prod_{j=1}^4(z-\beta_j)}} \rmd z=0,\\ \\
\oint_{\Gamma_{\infty}}z\,\frac{W'(z)}{\sqrt{\prod_{j=1}^4(z-\beta_j)}} \rmd z=0, \\ \\
\oint_{\Gamma_{\infty}}z^2\,\frac{W'(z)}{\sqrt{\prod_{j=1}^4(z-\beta_j)}} \rmd z=4\,\pi\,\rmi\,s,\\ \\
\oint_{\Gamma_{\infty}}z^3\,\frac{W'(z)}{\sqrt{\prod_{j=1}^4(z-\beta_j)}} \rmd z
=4\,\pi\,\rmi\,\Big(\frac{1}{2}\sum_{j=1}^4 \beta_j-N\,g_N\Big)\,s-\pi\,\rmi\, t_2.
\end{array}\right.
\end{equation}

\subsection{The classes $\mathcal{R}_q$ of solutions and their singular sectors}

The reduced systems allow us to establish a simple  characterization  of  relevant subsets of singular spectral curves.

\begin{defi} We define  $\mathcal{R}_q$ as the set of all  the solutions $(\boldsymbol\beta, \emph{\bt}^{(q)})$ of the reduced system (\ref{criti0})
such that $\beta_i\neq \beta_j$  for all $i\neq j$
\end{defi}
Obviously for spectral curves in the regular sector ${\rm reg}\mathcal{M}_q$ we have that $(\boldsymbol\beta(\bt^{(q)}), \bt^{(q)})\in
\mathcal{R}_q $.
Nevertheless, we are interested in degenerate solutions $(\boldsymbol\beta,\bt^{(q)})\in\mathcal{R}_q $  at which the Jacobian of the system
(\ref{criti0}) vanishes.
\begin{equation} \label{siei}
\Big |\frac{\partial ^{2}\mathbb{W}(\boldsymbol\beta,\bt)}{
\partial \beta_{i}\,\partial \beta_{j}}\Big|=0.
\end{equation}
From the Euler-Poisson-Darboux equations (\ref{eulp}) it follows  that  if
$(\boldsymbol\beta,\bt^{(q)})\in\mathcal{R}_q $  then  all the derivatives of $\mathbb{W}$  at $(\boldsymbol\beta,\bt^{(q)})$ can be expressed in
terms of linear combinations of derivatives of the
form $\partial _{\beta_{i}}^{k}\mathbb{W}$. In particular if $(\boldsymbol\beta,\bt^{(q)})\in\mathcal{R}_q $ we have
\[
\Big |\frac{\partial ^{2}\mathbb{W}(\boldsymbol\beta,\bt)}{
\partial \beta_{i}\,\partial \beta_{j}}\Big|=\prod_{i=1}^{2q}\frac{
\partial ^{2}\mathbb{W}(\boldsymbol\beta,\bt)}{\partial \beta_{i}^{2}}
.
\]
 Thus, it is natural to  define singular sectors in $\mathcal{R}_q$
as follows

\begin{defi}  Given a set of $2q$ non negative integers ${\bf n}=(n_1,n_2,\cdots,n_{2q})$ with some $n_i\geq 1$,
the singular sector ${\rm sing}\mathcal{R}_q({\bf n})$  is defined  as the set of points $(\boldsymbol\beta,\emph{\bt}^{(q)})\in \mathcal{R}_q$
such that
\begin{equation}\everymath{\displaystyle}\label{m12}
\left\{\begin{array}{l}  \frac{\partial ^{k}\mathbb{W}}{\partial \beta_{i}^{k}
}=0 ,\quad   \forall 1\leq k\leq n_{i}+1,  \\Ê\\
\frac{\partial ^{n_{i}+2}\mathbb{W}}{\partial \beta_{i}^{n_{i}+2}}\neq 0
, \end{array}\right.
\end{equation}
for  $i=1,2,\ldots
,2q$.
\end{defi}
\begin{theo} If a spectral curve  in ${\rm sing}\mathcal{M}_q$ has roots  $\beta_i$ with  multiplicities $2n_i+1$, then
$(\boldsymbol\beta(\emph{\bt}^{(q)}), \emph{\bt}^{(q)})\in{\rm sing}\mathcal{R}_q({\bf n})$ .
\end{theo}

\noindent
{\bf Proof}
From  (\ref{susper0}) and (\ref{yw})  it is clear that  at $\boldsymbol\beta=\boldsymbol\beta(\bt^{(q)})$   each  derivative $\partial_{\beta_i}^{k+1}
\mathbb{W} $ is proportional to the integral
\begin{eqnarray}\label{inted}
\nonumber \oint_{\Gamma_{\infty}}\frac{y_{0}(z,\boldsymbol\beta)\,y(z,{\bf t})}{(z-\beta_i)^{k+1}}\,\frac{\rmd z}{2\,\pi \,\rmi}=
\oint_{\Gamma_{\infty}}\frac{\prod_{l=1}^p (z-\alpha_l)\,\prod_{j\neq i} (z-\beta_j)}{(z-\beta_i)^{k}}\,\frac{\rmd z}{2\,\pi \,\rmi} \\ \\
\nonumber =\frac{1}{(k-1)!}
\partial_z^{k-1} \Big(\prod_{l=1}^p (z-\alpha_l)\,\prod_{j\neq i} (z-\beta_j)  \Big)\Big|_{z=\beta_i}.
\end{eqnarray}
Then the statement follows since,  according to our assumption,  the function $\prod_{l=1}^p (z-\alpha_l)$  has  zeros of order $n_i$ at $z=\beta_i$.

As a consequence of this theorem we conclude that the sets ${\rm sing}\mathcal{R}_q({\bf n})$ represent singular spectral curves in which
double zeros $\alpha_l$ of $y(z)^2$ coalesce with  simple zeros $\beta_i$. Figure 4  exhibits  a process of this type in the cubic model.

\section{ Critical behavior and the Multiple scaling limit method }

\bigskip

The observation made in the previous section drastically simplify the
analysis of the singular behavior of spectral curves  near
branch points of multiplicity higher than one, as well as
its possible regularization.

Let us consider a spectral curve such that at a certain value $\bt^{(q)}_0$ the roots  $\beta_{i0}$ have  multiplicities $2n_i+1$, then  at
$(\boldsymbol\beta(\bt^{(q)}_0), \bt^{(q)}_0)$
we have
\begin{equation}
\quad \frac{\partial ^{k}\mathbb{W}}{\partial \beta _{i}^{k}}=0,\quad
\forall 1\leq k\leq n_{i}+1,\quad \frac{\partial ^{n_{i}+2}\mathbb{W}}{
\partial \beta _{i}^{n_{i}+2}}\neq 0 \label{7.1}
\end{equation}
for $i=1,\ldots ,2q$,   and in virtue of  (\ref{eulp}) we also have
\begin{equation}
\frac{\partial ^{l_{1}+...+l_{2 q}}\mathbb{W}}{\partial \beta
_{1}^{l_{1}}...\partial \beta _{2q}^{l_{2q}}}=0,\qquad 1\leq l_i \leq n_i+1.
\label{7.2}
\end{equation}
To
analyze the behavior of  $\boldsymbol\beta(\bt^{(q)}_0)$ near  $
\bt^{(q)}_0=(t_{01},\ldots,t_{0q})$  we will use the standard technique of the multiple scaling
limit (see e.g. ~\cite{DI95}). So, we take

\begin{equation}
t_{k}=t_{0k}+\varepsilon ^{\alpha }t_{k}^{\ast },\qquad \beta _{i}=\beta
_{i0}+\varepsilon ^{\gamma _{i}}\beta _{i}^{\ast },
\label{7.3}
\end{equation}
where $\varepsilon $ $\ll 1$ and $\alpha $ and $\gamma _{i}$ are positive
parameters to be determined. Now we expand $\mathbb{W}(\boldsymbol\beta ,\bt^{(q)})$ near the
point $(\boldsymbol\beta _{0},\bt^{(q)}_0)$. Writing $\mathbb{W}(\boldsymbol\beta ,\bt^{(q)})=\mathbb{W}^{(0)}
(\boldsymbol\beta)+\sum_{k=1}^{q}t_{k}\mathbb{V}_{k}(\boldsymbol\beta)$, taking into
account (\ref{7.1}), (\ref{7.2}) and keeping the leading terms in the expansion,
one gets

\begin{eqnarray*}
\mathbb{W}(\boldsymbol\beta ,\bt^{(q)})=\mathbb{W}_{0}+\varepsilon
^{\alpha }\sum_{k=1}^{q}t_{k}^{\ast }A_{k}+\varepsilon ^{\alpha
}\sum_{k=1}^q\sum_{l=1}^{2q}\varepsilon ^{\gamma _{l}}t_{k}^{\ast }A_{kl}\beta
_{l}^{\ast }\\
+\varepsilon ^{\alpha }\sum_{k=1}^q\sum_{l,m=1}^{2q}\varepsilon ^{\gamma
_{l}+\gamma _{m}}t_{k}^{\ast }A_{klm}\beta _{l}^{\ast }\beta _{m}^{\ast
}+\sum_{k=1}^{2q}\varepsilon ^{(n_{k}+2)\gamma _{k}}B_{k}\,(\beta _{k}^{\ast
})^{n_{k}+2}+...  \label{7.4}
\end{eqnarray*}
where
\[
\mathbb{W}_{0}=\mathbb{W}(\boldsymbol\beta _{0},\bt^{(q)}_0),\quad A_{k}=\mathbb{V}_{k}(\boldsymbol\beta _{0}),\quad A_{kl}=\frac{\partial \mathbb{V}
_{k}}{\partial \beta _{l}}(\boldsymbol\beta _{0}),
\]
\[
A_{klm}=\frac{1}{2}\frac{
\partial \mathbb{V}_{k}}{\partial \beta _{l}\partial \beta _{m}}(
\boldsymbol\beta _{0}),\quad B_{k}=\frac{1}{(n_{k}+2)!}\frac{\partial ^{n_{k}+2}\mathbb{W}
}{\partial \beta _{k}^{n_{k}+2}}(\boldsymbol\beta _{0},\bt^{(q)}_0).
\]
 Consequently,
\begin{eqnarray}
\nonumber
\frac{\partial \mathbb{W}}{\partial \beta _{l}^{\ast }}=\varepsilon
^{\alpha +\gamma _{l}}\sum_{k=1}^{q}t_{k}^{\ast }A_{kl}+\varepsilon ^{\alpha
+\gamma _{l}}2\sum_{k=1}^q\sum_{m=1}^{2q}\varepsilon ^{\gamma _{m}}t_{k}^{\ast
}A_{klm}\beta _{m}^{\ast }\\
\label{7.5}
\\
\nonumber
+\varepsilon ^{(n_{l}+2)\gamma
_{l}}(n_{l}+2)B_{l}\,(\beta _{l}^{\ast })^{n_{l}+1}+....
\end{eqnarray}
In the generic case when at least one  $A_{kl}$ does not vanish,  the second term is
subdominant with respect to the first one.  In such a case the balance between the first
and third terms is achieved if $\alpha +\gamma _{l}=(n_{l}+2)\gamma _{l},$
i.e.

\begin{equation}
\gamma_l=\frac{\alpha}{n_{l}+1},\qquad l=1,...,2q.  \label{7.7}
\end{equation}
Thus, in the generic case the expansion of $\mathbb{W}(\boldsymbol\beta ,\bt^{(q)})$ is of the form

\begin{equation}\label{7.7a}
\mathbb{W}(\boldsymbol\beta ,\bt^{(q)})=\mathbb{W}_{0}+\varepsilon
\sum_{k=1}^{q}t_{k}^{\ast }A_{k}
+ \sum_{l=1}^{2q}\varepsilon ^{
\frac{n_l+2}{n_{l}+1}}\Big(\xi _{l}\beta _{l}^{\ast }+B_{l}(\beta _{l}^{\ast
})^{n_{l}+2}\Big)+\ldots
\end{equation}
where $\xi _{l}=\sum_{k=1}^{q}t_{k}^{\ast }A_{kl}$ and we put $\alpha =1$.

Equations (\ref{7.5})-(\ref{7.7}) imply that when $\bt^{(q)}$ and $\boldsymbol\beta$ are approaching $
\bt^{(q)}_0$ and $\boldsymbol\beta_0$ then

\begin{equation} \label{7.8}
\beta _l-\beta _{l0}\sim \Big(\sum_{k=1}^{q}(t_{k}-t_{0k})A_{kl}\Big)^{\frac{1}{
n_l+1}},\qquad l=1,...,2q.
\end{equation}
and hence the derivatives of $\beta _l$ blow up as

\begin{equation} \label{7.9}
\frac{\partial \beta _l}{\partial t_{k}}\sim
\Big(\sum_{k=1}^{q}(t_{k}-t_{0k})A_{kl}\Big)^{-\frac{n_l}{n_l+1}},
\end{equation}
as $\bt^{(q)}\rightarrow \bt^{(q)}_0$.

An unbounded increase of derivatives of $\beta _l$ while approaching the
singular sectors is the origin of the break of analyticity for the
spectral curve and the free energy $(\ref{free})$. From the point of view of the hydrodynamic type equations associated to  the
differential equations (\ref{diff}), in particular for the Burgers-Hopf
equations
\begin{equation}
\frac{\partial \beta _{l}}{\partial t_{j}}=\beta _{l}^{N-j}\frac{\partial
\beta _{l}}{\partial t_{N}},\qquad l=1,..,N,\qquad j=1,..,N-1,  \label{7.10}
\end{equation}
associated with the equations (\ref{sn}), an unbounded increase of
derivatives of $\beta _{l}$ represents the classical gradient catastrophe (see e.g.  ~\cite{WH,RJ}).
Within the theory of polynomials with multiple roots and
singularities of $A_{N+1}$ type, this singular behavior   is the analytical manifestation of the
catastrophe arising in the transitions between two different strata (see e.g.
~\cite{A,AVG,TH}).

\section{Regularization of gradient catastrophes}
In all physical models where such catastrophe with derivatives happens it is
regularized at the end by one or another mechanism. It is, for instance,
dissipation or dispersion for hydrodynamical models (see e.g. (~\cite{WH})), quantum
corrections or higher genus expansion in matrix models (see e.g. (~\cite{DI95})).
Formally such mechanisms consist in the appearance of higher order
derivatives in the basic relations or differential equations of the type (\ref
{diff}) or (\ref{7.10}). Concrete forms of these higher order corrections
typically are obtained by appropriate expansion or limit for the known
model exact relations or equations.

Here we suggest an approach for regularization which
goes in the direction opposite to the standard ones. The idea is to deform
the scheme presented above in such model independent manner that it
naturally leads to the appearance of derivatives in equations (\ref{7.5}) and
(\ref{7.8}) to prevents the blow up of $\partial_{t_{k}} \beta _i$. The principal point is the substitution of the equations
(\ref{criti0}) or (\ref{7.7a}) for critical points of the functions $\mathbb{W}
$ by Euler-Lagrange equations
\begin{equation}\label{7.11}
\frac{\delta \mathbb{W}^{reg}}{\delta \beta _{i}}=0,  \quad i=1,\ldots,2q,
\end{equation}
 where $\mathbb{W}
^{reg}$ is an appropriate modification of $\mathbb{W}$ obtained by
adding terms with derivatives. Thus we rewrite (\ref{7.7a}), dropping
the insignificant second term, as

\begin{equation}\label{7.12}
\mathbb{W}(\boldsymbol\beta,\bt^{(q)})=\mathbb{W}_{0}+\varepsilon
\sum_{l=1}^{2q}\varepsilon ^{\frac{1}{n_{l}+1}}\mathbb{U}_{l}(\beta
_{l}^{\ast })+\cdots
\end{equation}
where $\mathbb{U}_{l}(\beta _{l}^{\ast })=\xi _{l}\beta _{l}^{\ast
}+B_{l}\,(\beta _{l}^{\ast })^{n_{l}+2}$.  Modification of the polynomial (\ref
{7.12}) is achieved by adding terms with derivatives of the functions $\beta_i $ of
appropriate order in $\varepsilon $. The general case is pretty involved and
will be studied in a separate paper. Here we will consider some specific
regularizations for particular  situations when the corresponding equations
(\ref{7.11}) are rather simple.

In the case when all $n_{l}$ are different the terms in  (\ref{7.12})
are of different degrees,  so it is natural to treat  each term  in the sum (\ref{7.12}) separately. The
simplest nontrivial modification  of $\mathbb{W}(\boldsymbol\beta,\bt^{(q)})$ is of the form

\[
\mathbb{W}^{reg}(\boldsymbol\beta^{\ast },\boldsymbol\xi)=\mathbb{W}
_{0}+
 \varepsilon\sum_{l=1}^{2q}\Big(\varepsilon ^{\frac{1}{
n_{l}+1}}\mathbb{U}_{l}(\beta _{l}^{\ast })
+\varepsilon ^{\delta _{l}}C_{l}\Big(
\frac{\partial \beta _{l}^{\ast }}{\partial \xi _{l}}\Big)^{2}\Big)+\cdots
\]
with some constants $C_{l}$ and appropriate $\delta _{l}$. A balance of
degrees in the Euler-Lagrange equation  (\ref{7.11}) or, equivalently, in the equation
\[
\frac{
\delta \mathbb{W}^{reg}}{\delta \beta _{i}^{\ast }}=\frac{\partial
\mathbb{W}^{reg}}{\partial \beta _{i}^{\ast }}-\sum_{k}\frac{\partial }{
\partial \xi _{k}}\Big(\frac{\partial \mathbb{W}^{reg}}{\partial(\partial_{\xi _{k}} \beta
_{i}^{\ast })}\Big)=0,
\]
 implies that
 \[
 \delta _{l}=\frac{1}{
n_{l}+1}.
\]
 Thus,
\begin{equation} \label{7.14}
\nonumber \mathbb{W}^{reg}(\boldsymbol\beta^{\ast },\boldsymbol \xi)=\mathbb{W}_{0} +
\sum_{l=1}^{2q} \varepsilon ^{\frac{n_l+2}{n_{l}+1}}\Big(\xi _{l}\beta _{l}^{\ast
}+B_{l}\,(\beta _{l}^{\ast })^{n_{l}+2}+C_{l} (\partial_{\xi _{l}} \beta _{l}^{\ast })^2
\Big)+\cdots
\end{equation}
and the Euler-Lagrange equations take the form

\begin{equation}\label{7.15}
2 C_{l}\frac{\partial ^{2}\beta _{l}^{\ast }}{\partial \xi _{l}^{2}}=\xi
_{l}+(n_{l}+2)B_{l}(\beta _{l}^{\ast })^{n_{l}+1},\quad l=1,...,2q.
\end{equation}

In the case when $n_{l}=1$ for one of the equations (\ref{7.15}) then  the simple redefinitions
\[
\beta _{l}^{\ast }=\Big(
\frac{16 C_{l}}{B_{l}^{3}}\Big)^{\frac{1}{5}}\Omega,\quad \xi _{l}=\Big(\frac{
8C_{l}^{2}}{B_{l}}\Big)^{\frac{1}{5}}x
\]
 convert this $l$-th equation into the Painlev\`{e}-I equation
\begin{equation}
\frac{\partial ^{2}\Omega }{\partial x^{2}}=6\Omega ^{2}+x.  \label{7.16}
\end{equation}
This equation has appeared as the regularizing equation
of different types of catastrophes (see e.g. ~\cite{DI95,DUB}). In our case
the Tritronqu\`{e}e solution of the Painlev\`{e}-I equation (\ref{7.16}) describes
the regularization of the blow up (\ref{7.9}) with $n_{l}=1$, and the analyticity
breaking associated with the formation of a third-order root from the merging of a double root $\alpha$ and
the simple
root $\beta _l$. A similar situation takes place for $n_{l}\geq 2$.

In the case when some of $n_{i}$ coincide, say, $n_{1}=n_{2}=...=n_{k}=n$
the polynomial $\mathbb{W}$ is

\begin{eqnarray}\label{7.17}
\nonumber \mathbb{W}(\boldsymbol\beta,\bt^{(q)})=\mathbb{W}_{0}+\varepsilon ^{
\frac{n+2}{n+1}}\sum_{l=1}^{k}\Big(\xi _{l}\beta _{l}^{\ast }+B_{l}(\beta
_{l}^{\ast })^{n+2}\Big)\\Ê\\
\nonumber +\varepsilon \sum_{l=k+1}^{2q}\varepsilon ^{\frac{1}{
n_{l}+1}}\Big(\xi _{l}\beta _{l}^{\ast }+B_{l}(\beta _{l}^{\ast })^{n_{l}+2}\Big)+\cdots
\end{eqnarray}
The dominant contribution of $\beta _{1}^{\ast },\beta _{2}^{\ast
},...,\beta _{k}^{\ast }$ into $\mathbb{W}$ is of the same
order. So, $\mathbb{W}^{reg}(\boldsymbol\beta^{\ast },\boldsymbol\xi )$ should naturally contains a mixture
of derivatives of $\beta _{1}^{\ast },\beta _{2}^{\ast },...,\beta
_{k}^{\ast }$. Hence, the simplest natural ansatz for the modified $\mathbb{W
}$ of (\ref{7.17}) would be

\begin{eqnarray}\label{7.18}
\nonumber \mathbb{W}^{reg}(\boldsymbol\beta^{\ast },\boldsymbol\xi) =\mathbb{W}_{0}+\varepsilon
^{\frac{n+2}{n+1}}\Big[ \sum_{l=1}^{k}\Big(\xi _{l}\beta _{l}^{\ast
}+B_{l}(\beta _{l}^{\ast })^{n+2}\Big)\\\nonumber \\+\frac{1}{2}\sum_{m,p,q=1}^{k}D_{mpq}\beta
_{m}^{\ast }\frac{\partial \beta _{p}^{\ast }}{\partial \xi _{q}}
 +\frac{1}{2}
\sum_{m,p,q,t=1}^{k}D_{mpqt}\frac{\partial \beta _{m}^{\ast }}{\partial \xi
_{p}}\frac{\partial \beta _{q}^{\ast }}{\partial \xi _{t}}\Big]\\\nonumberÊ\\
\nonumber
+\varepsilon \sum_{l=k+1}^{2q} \varepsilon ^{\frac{1}{n_{l}+1}}\Big(\xi
_{l}\beta _{l}^{\ast }+B_{l}(\beta _{l}^{\ast })^{n_{l}+2}+C_{l}\Big(\frac{
\partial \beta _{l}^{\ast }}{\partial \xi _{l}}\Big)^{2}\Big)+\cdots
\end{eqnarray}
where $C_{l},D_{mpq},D_{mpqt}$ are some constants  such that $
D_{mpq}=-D_{pmq},D_{mpqt}=D_{qtmp}$. The Euler-Lagrange equations corresponding to the
variables $\beta _1^{\ast },\ldots,\beta _k^{\ast }$ are

\begin{equation}\label{7.19}
\sum_{m,p,q=1}^{k}D_{impq}\frac{\partial ^{2}\beta _{p}^{\ast }}{\partial
\xi _{m}\partial \xi _{q}}-\sum_{m,p=1}^{k}D_{imp}\frac{\partial \beta
_{m}^{\ast }}{\partial \xi _{p}}=\xi _{i}+(n+2)B_{i}(\beta _{i}^{\ast
})^{n+1},
\end{equation}
while those corresponding to the variables $\beta _{k+1}^{\ast },...,\beta _{s}$ are the
equations (\ref{7.15}).

In the simplest case $k=2$ and under the constraints $D_{impq}=0$, the system (\ref{7.19})
is equivalent to

\begin{eqnarray}
\nonumber \frac{\partial \beta _{1}^{\ast }}{\partial \xi } &=&D_{122}\xi +C\eta
+(n+2)B_{2}\beta _{2}^{\ast n+1}, \\ \label{7.191}\\
\nonumber \frac{\partial \beta _{2}^{\ast }}{\partial \xi } &=&-D_{121}\xi -A\eta
-(n+2)B_{1}\beta _{1}^{\ast n+1} ,
\end{eqnarray}
where the variables $\xi $ and $\eta $ are defined as $\xi _{1}=D_{121}\xi
+A\,\eta ,\xi _{2}=D_{122}\xi +C\,\eta $,  with  $A$  and $C$ being arbitrary constants.  For $ n=1$
and $A=C=0$  it is the special Riccati system.  For  $n=2$ and $A=C=0$ it is a
particular instance of the differential cubic systems which can be viewed as
the multi-component generalizations of the Abel equation (see e.g. ~\cite{MU}).

At $k=2$ and under the constraints $D_{mpq}=0$, with the change of variables $\xi _{1}=\xi+\eta$
$\xi_{2}=\xi-\eta$ and restricted to the subspace $\eta=0$ the system (\ref{7.19}) takes the form
\begin{eqnarray}
\nonumber \frac{\partial ^{2}\beta _{1}^{\ast }}{\partial \xi ^{2}} &=&A_{11}\beta
_{1}^{\ast n+1}+A_{12}\beta _{2}^{\ast n+1}+A_{1}\xi , \\ \label{7.20} \\
\frac{\partial ^{2}\beta _{2}^{\ast }}{\partial \xi ^{2}} &=&A_{21}\beta
_{1}^{\ast n+1}+A_{22}\beta _{2}^{\ast n+1}+A_{2}\xi ,  \nonumber
\end{eqnarray}
where
$$\everymath{\displaystyle}\begin{array}{l}
A_{11}=\frac{4(n+2)B_1B_{22}}{B_{11}B_{22}-B_{12}^2},\quad
A_{12}=-\frac{4(n+2)B_2B_{12}}{B_{11}B_{22}-B_{12}^2},\quad
A_1=\frac{4(B_{22}-B_{12})}{B_{11}B_{22}-B_{12}^2},\\  \\
A_{21}=-\frac{4(n+2)B_1B_{12}}{B_{11}B_{22}-B_{12}^2},\quad
A_{22}=\frac{4(n+2)B_2B_{11}}{B_{11}B_{22}-B_{12}^2},\quad
A_2=\frac{4(B_{11}-B_{12})}{B_{11}B_{22}-B_{12}^2},\\  \\
B_{lm}=\displaystyle\sum_{i,j=1,2}D_{limj}.
\end{array}$$
The system (\ref{7.19}) for arbitrary $k$ also admits special reductions under
which it takes  a form similar to (\ref{7.20}). The systems (\ref{7.19})-(\ref
{7.20}) describe  regularizations of particular catastrophes associated
with the singular sectors $\mathrm{sing}\mathcal{R}({\bf n})$ such that $n_1=\ldots=n_k=n$. Detailed study of the properties of
such systems and  their implications for the string and matrix theory will
be presented in a separate publication.

\ack
L. Mart\'{i}nez Alonso and E. Medina  are grateful to G. \'Alvarez for many useful conversations on the subject of spectral curves
in gauge/string dualities.
The financial support of the Universidad Complutense under project GR58/08-910556, the Comisi\'on Interministerial de Ciencia y
Tecnolog\'{\i}a under project  FIS2011-22566  and PRIN 2008 grant no. 28002K9KXZ are gratefully acknowledged.
\section*{Appendix A}
In this appendix we briefly discuss the elements of the  theory of Abelian differentials in Riemann surfaces that we use
in section 2.

Let $M$ be the genus $q-1$ Riemann surface associated to (\ref{1.1}), it is determined by two sheets
\begin{equation}
    M = M_1\cup M_2,\quad M_i = \{Q=(y_{0,i}(z),z)\},
\end{equation}
where
\[
y_{0,2}(z)=-y_{0,1}(z),\quad y_{0,1}(z)\sim z^q,\quad z\rightarrow\infty.
\]
We  introduce the following Abelian differentials in $M$:
 \begin{description}
    \item[(1)] The canonical basis of first kind (i.e., holomorphic) Abelian differentials  $\{ \rmd \varphi_i\}_{i=1}^{q-1}$
            with the normalization $A_i(\rmd \varphi_j )=\delta_{ij}$. These differentials can be written as
            \begin{equation}
                \label{nho}
                \rmd \varphi_j(z) = \frac{p_j(z)}{y_0(z)}\rmd z,
            \end{equation}
            where the $p_j(z)$ are polynomials of degree not greater than $q-2$ uniquely determined
            by the normalization conditions.
    \item[(2)] The second kind Abelian differentials $\rmd \Omega_n$  $(n\geq 1)$ whose only poles are at
            $\infty_1$, such that
            \begin{equation}
                \label{ok}
                \rmd \Omega_n(Q)
                =
                (n z^{n-1}+\mathcal{O}(z^{-2}))\rmd z,\quad Q\rightarrow \infty_1,\quad z=z(Q),
            \end{equation}
            and normalization $A_i(\rmd \Omega_n)=0$ $(i=1,\ldots,q-1)$. It is easy to see that
            \begin{equation}
                \label{diff1}
                \rmd \Omega_n=\left(\frac{n}{2} z^{n-1}+\frac{P_n(z)}{y_{0}(z)}\right)\rmd z,
            \end{equation}
            where the $P_n(z)$ are polynomials of the form
            \begin{equation}
                \label{ek}
                P_n(z) = \frac{n}{2} (z^{n-1} y_0(z))_{\oplus}+\sum_{i=0}^{q-2}c_{n i} z^i
            \end{equation}
            and the coefficients $c_{n i}$ are uniquely determined by the normalization conditions.
    \item[(3)] The third kind Abelian differential  $\rmd\Omega_0$ whose only poles are at $\infty_1$ and $\infty_2$, such that
            \begin{equation}
                \label{o0}
                \rmd \Omega_0(Q)
                =
                \left\{\begin{array}{ll}
                                        \displaystyle\left(\frac{1}{z}+\mathcal{O}(z^{-2})\right)\rmd z,&  Q\rightarrow \infty_1\\\\
                                       \displaystyle\left(-\frac{1}{z}+\mathcal{O}(z^{-2})\right)\rmd z,& Q\rightarrow \infty_2,
                                \end{array}\right.
            \end{equation}
            and normalization $A_i(\rmd \Omega_0)=0$ for all $i=1,\ldots,q-1$. It follows that
            \begin{equation}
                \label{diff10}
                \rmd\Omega_0 = \frac{P_0(z)}{y_{0}(z)} \rmd z,
            \end{equation}
            where $P_0(z)$ is a polynomial of the form
            \begin{equation}
                \label{e0}
                P_0(z)=(z^{-1} y_{0}(z))_{\oplus}+\sum_{i=0}^{q-2}c_{0 i} z^i
            \end{equation}
            and the coefficients $c_{0 i}$ are uniquely determined by the normalization conditions.
\end{description}
We notice that all the polynomials $p_j(z)$ and $P_n(z)$ also depend on the set of $2q$ branch points  $\boldsymbol\beta$,
but for simplicity this dependence will not be indicated unless necessary.


%
\section*{References}
\bibliographystyle{iopart-num}
\bibliography{hmms}

\providecommand{\newblock}{}
\begin{thebibliography}{10}
\expandafter\ifx\csname url\endcsname\relax
  \def\url#1{{\tt #1}}\fi
\expandafter\ifx\csname urlprefix\endcsname\relax\def\urlprefix{URL }\fi
\providecommand{\eprint}[2][]{\url{#2}}

\bibitem{CA01}
Cachazo F, Intriligator K and Vafa C 2001 {\em Nuc.\ Phys.\ B\/} {\bf 603} 3

\bibitem{DI02}
Dijkgraaf R and Vafa C 2002 {\em Nuc.\ Phys.\ B\/} {\bf 644} 3

\bibitem{DI022}
Dijkgraaf R and Vafa C 2002 {\em Nuc.\ Phys.\ B\/} {\bf 644} 21

\bibitem{SE94}
Seiberg N and Witten E 1994 {\em Nuc.\ Phys.\ B\/} {\bf 426} 19

\bibitem{CA02b}
Cachazo F and Vafa C 2002 {\em arXiv:hep-th/020601\/}

\bibitem{DE99}
Deift P 1999 {\em Orthogonal Polynomials and Random Matrices: A
  {Riemann--Hilbert} approach\/} (Providence: American Mathematical Society)

\bibitem{BL08}
Bleher P 2008 {\em Lectures on random matrix models. The {Riemann-Hilbert}
  approach\/} (Amsterdam: North Holland)

\bibitem{AL10}
\'Alvarez G, {Mart\'{\i}nez Alonso} L and Medina E 2010 {\em J. Stat.\ Mech.\
  Theory Exp.\/}  03023

\bibitem{LA03}
Lazaroiu C~I 2003 {\em J. High Energy Phys.\/} {\bf 03} 044

\bibitem{FE04}
Felder G and Riser R 2004 {\em Nuc.\ Phys.\ B\/} {\bf 691} 251

\bibitem{BI05}
Bilal A and Metzger S 2005 {\em J. High Energy Phys.\/} {\bf 08} 097

\bibitem{GO87}
Gonchar A and Rakhmanov E~A 1987 {\em Math.\ {USSR} Sbornik\/} {\bf 62}

\bibitem{GO89}
Gonchar A~A and Rakhmanov E~A 1989 {\em Math.\ {USSR} Sbornik\/} {\bf 62} 305

\bibitem{BL99}
Bleher P and Its A 1999 {\em Ann.\ Math.\/} {\bf 150} 185--266

\bibitem{BE06}
Bertola M, Eynard B and Harnad J 2006 {\em Commun.\ Math.\ Phys.\/} {\bf 263}
  401

\bibitem{BE09}
Bertola M and Mo M~Y 2009 {\em Adv.\ Math.\/} {\bf 220} 154

\bibitem{BE11}
Bertola M 2011 {\em Analysis and Math.\ Phys.\/} {\bf 1} 167

\bibitem{MA11}
Mart{\'{\i}}nez-Finkelshtein A and Rakhmanov E~A 2011 {\em Commun.\ Math.\
  Phys.\/} {\bf 302} 53

\bibitem{RA12}
Rakhmanov E~A 2012 Orthogonal polynonials and s-curves {\em Recent advances in
  orthogonal polynomials, special functions and their applications\/} vol 578
  of Contemp. Math. (Amer. Math. Soc. Providence, RI.) pp 195--239

\bibitem{AL13}
\'Alvarez G, {Mart\'{\i}nez Alonso} L and Medina E 2013 {\em J. High Energy
  Phys.\/} {\bf 03} 170

\bibitem{FE03d}
Ferrari F 2003 {\em Phys.\ Rev.\ D\/} {\bf 67} 085013

\bibitem{KR04}
Krichever I 1994 {\em Comm.Pure Appl. Math.\/} {\bf 47} 437

\bibitem{ch03}
Chekhov L and Mironov A 2003 {\em Phys.\ Lett.\ B\/} {\bf 552} 293

\bibitem{ch03b}
Chekhov L, Marshakov A, Mironov A and Vasiliev D 2003 {\em Phy.\ Lett.\ B\/}
  {\bf 562} 323

\bibitem{DAR}
Darboux G 1915 {\em Lecons sur la th\`{e}orie g\`{e}n\`{e}rale des surfaces II
  surfaces II,\/} (Gauthier Villars)

\bibitem{DI95}
{Di Francesco} P, Ginsparg P and Zinn-Justin J 1995 {\em Phys.\ Rep.\/} {\bf
  254} 1--133

\bibitem{KO11}
Konopelchenko B, {Mart\'{\i}nez Alonso} L and Medina E 2011 {\em Phys.\ Lett.\
  A\/} {\bf 375} 867--872

\bibitem{MAR10}
Mari{\~n}o M, Pasquetti S and Putrov P 2010 {\em J.\ High Energy Phys.\/} {\bf
  10} 074

\bibitem{BE12}
Bertola M and Tovbis A 2011 {\em {arXiv:1108.0321}\/}

\bibitem{GI82}
Givental A~B 1982 {\em Functional Analysis and its Applications\/} {\bf 16}
  10--14

\bibitem{WH}
Whitham G~B 1974 {\em Linear and nonlinear waves\/} (J.Wiley and Sons,
  New-york)

\bibitem{RJ}
Rozdestvenski B and Yanenko N 1983 {\em Systems of quasilinear 
  their applications to gas dynamics\/} (Math.Monog., v. 55, AMS, Providence,
  RI.)

\bibitem{A}
Arnold V~I 1976 {\em Comm.Pure Appl. Math.\/} {\bf 29} 557--582

\bibitem{AVG}
Arnold V~I, Varchenko A~N and Guseyn-Zade S~M 1985 {\em Singularities of
  differentiable maps\/} (Birkhauser, Boston, Ma.)

\bibitem{TH}
Thom R 1989 {\em Structural stability and morphogenesis\/} (Addison-Wesley,
  P.C. ,CA)

\bibitem{DUB}
Dubrovin B 2008 On universality of critical behaviour in hamiltonian pdes {\em
  Geometry, topology and mathematical physics\/} (AMS Transl., 224, Ser. 2,
  59-109, AMS, Providence, RI)

\bibitem{MU}
Murphy G 1960 {\em Ordinary differential equations and their solutions\/} (Van
  Vostrand,)

\end{thebibliography}
\end{document}